\documentclass[10pt]{article}
\usepackage[margin=1in]{geometry}
\usepackage{hyperref}
\usepackage{multicol}
\usepackage{amsmath}
\usepackage{amssymb}
\usepackage{mathrsfs}
\usepackage{dsfont}
\usepackage{framed}
\usepackage{graphicx}
\usepackage[table]{xcolor}
\usepackage{framed,color}
\definecolor{shadecolor}{rgb}{0.85,0.85,0.85}
\usepackage{url}
\usepackage{cleveref}
\usepackage{authblk}

\title{The Network HHD: Quantifying Cyclic Competition in Trait-Performance Models of Tournaments}
\author[1]{Alexander Strang}
\author[2]{Karen C. Abbott}
\author[3]{Peter J. Thomas}

\affil[1]{Department of Statistics, University of Chicago}
\affil[2]{Department of Biology, Case Western Reserve University}
\affil[3]{Department of Mathematics, Case Western Reserve University}

\date{\today}

\begin{document}

\maketitle

\section{Abstract}
 Competitive tournaments appear in sports, politics, population ecology, and animal behavior.  All of these fields have developed methods for rating competitors and ranking them accordingly.  A tournament is intransitive if it is not consistent with any ranking.  Intransitive tournaments contain rock-paper-scissor type cycles.  The discrete Helmholtz-Hodge decomposition (HHD) is well adapted to describing intransitive tournaments.  It separates a tournament into perfectly transitive and perfectly cyclic components, where the perfectly transitive component is associated with a set of ratings. The size of the cyclic component can be used as a measure of intransitivity.  Here we show that the HHD arises naturally from two classes of tournaments with simple statistical interpretations.  We then discuss six different sets of assumptions that define equivalent decompositions. This analysis motivates the choice to use the HHD among other existing methods. Success in competition is typically mediated by the traits of the competitors. A trait-performance model assumes that the probability that one competitor beats another can be expressed as a function of their traits. We show that, if the traits of each competitor are drawn independently and identically from a trait distribution then the expected degree of intransitivity in the network can be computed explicitly. Using this result we show that increasing the number of pairs of competitors who could compete promotes cyclic competition, and that increasing the correlation in the performance of $A$ against $B$ with the performance of $A$ against $C$ promotes transitive competition. The expected size of cyclic competition can thus be understood by analyzing this correlation. An illustrative example is provided.
 
 	\section{Introduction: Tournaments, Ranking, and Intransitivity} \label{sec: Intro}
	
	Competitive tournaments are important across disciplines. Examples range from ecology and animal behavior \cite{Laird, Shizuka}, to psychology and sports \cite{Bozoki,Kendall}. Rating and ranking is important in each of these areas. In sports, ranking and rating teams and players is a topic of broad popular interest. In biology, fitness is an intrinsic rating since survival and reproduction are influenced by repeated competitive interactions with many individuals. Ranking is especially important in politics, as many electoral systems determine a winner by aggregating votes into a partial ranking of the candidates. Ratings and rankings are often sought since they simplify the description of a tournament by assigning each competitor a single number that purports to measure how good they are.
	
	Not all tournaments allow for a consistent ranking of competitors. This observation motivates classification into transitive and intransitive tournaments. A tournament is \emph{transitive} if knowing that $A$ usually beats $B$, and $B$ usually beats $C$, is enough to conclude that $A$ usually beats $C$. Transitive tournaments are consistent with a global ranking of all the competitors. 
	An \emph{intransitive} tournament is a tournament that is not consistent with any global ranking. 
	Intransitive tournaments must contain at least one cycle where the transitive assumption fails. Examples of intransitive tournaments appear in practically every discipline where tournaments are studied \cite{Candogan,Gehrlein,May,Petraitis,Reichenbach_a}, and are the norm rather than the exception when using real data \cite{Jiang,Kendall,Kerr,Laird,Shizuka,Sinervo,Slater}. Intransitivity may arise due to uncertainty in observed data \cite{Kendall,Slater}, or may be intrinsic to competition as in the game of rock-paper-scissors.
	
	Intransitivity is important for two reasons.
	
	First, intransitivity presents a challenge when ranking competitors since no ranking is consistent with the tournament. For example, Condorcet's paradox is a voting paradox in which voter's preferences lead to cyclic community preferences \cite{Gehrlein}.\footnote{Suppose there are three candidates in an election and three voters. Suppose that the first voter prefers A to B to C, the second B to C to A, and the third C to A to B. Then A would beat B in an election between the pair, B would beat C, and C would beat A.} Because of the cyclic community preferences there is no way to fairly rank the candidates, and, as a consequence, pick a winner of the election. 
	    
	Second, when intransitivity is intrinsic to the structure of the tournament then the tournament contains cyclic structure, as in rock-paper-scissors. Cyclic structures can radically alter optimal strategies \cite{Candogan} and long term dynamics \cite{May,Reichenbach_a,Reichenbach_b,Reichenbach_c,Reichenbach_d}. For example, in ecology it is widely hypothesized that intransitive competition between species promotes biodiversity since no species dominates. This hypothesis is based on extensive theoretical work \cite{Laird,May,Reichenbach_a,Reichenbach_b,Reichenbach_c,Reichenbach_d,Xue} and limited case-studies of small species assemblages \cite{Jackson,Kerr,Lankau_a,Lankau_b,Sinervo}. However, the importance of intransitivity in real natural communities is controversial \cite{Godoy,Soliveres,Ulrich} - in part because there are few robust metrics for measuring intransitivity from incomplete and noisy data. It has been shown that uncertainty in data can easily be conflated with observed intransitivity, and that common sampling methods for filling in missing data overestimate intransitivity \cite{Shizuka}.
	
    Thus there is a need for ranking and rating methods that are robust to intransitivity and measures of intransitivity that can handle noisy and incomplete data.
	
	Jiang et al introduced the discrete Helmholtz-Hodge Decomposition (HHD) as a general method for ranking objects from incomplete and imbalanced data \cite{Jiang}. 
	The decomposition is a network theoretic tool that we adapt to the study of competitive tournaments. 
	The HHD accomplishes three fundamental tasks. First, it assigns a rating to each competitor. Competitors can be ranked accordingly. Second, it produces a measure of intransitivity that quantifies how far an observed network is from the nearest perfectly transitive network. Third, it represents the observed network as the direct sum of perfectly transitive and a perfectly cyclic networks. This decomposition provides an elegant characterization of intransitivities present in data, and can reveal underlying cyclic tendencies in tournaments. This last property was leveraged by Candogan to identify cyclic structures within collections of competing strategies \cite{Candogan}. 
	
	When compared to existing ranking methods and intransitivity measures, the discrete HHD is attractive has a number of advantages. It is more general than some classical methods since it applies to arbitrary network topologies and can accommodate imbalanced data \cite{Jiang}. It is also more informative because it provides a clear description of both underlying transitive and cyclic structures. Most ranking methods and intransitivity measures focus on the transitive component while the HHD puts the transitive and cyclic components on equal footing. Finally, it remains efficiently computable even for large, incomplete networks \cite{Jiang}. 
	In contrast, Slater's index \cite{Slater} requires solving an NP hard optimization problem \cite{Charbit,Eades}, and Kendall's index \cite{Kendall} requires a complete network.
	
	This paper aims to answer two fundamental questions:
	
	\begin{enumerate}
	    \item[\textbf{1.}]  Why use the HHD when other methods exist?
	    \item[\textbf{2.}]  Having chosen to use the HHD, what do we expect when pairwise competitive advantage derives from traits drawn from an underlying distribution?
	\end{enumerate}
	
	Answering the first question is important since there are many possible methods to choose from, so the choice of method should be made in a principled way. Answering the second question is important since it builds a conceptual bridge from the competitors and competitive event to the overall structure of tournament. As in Landau \cite{Landau_a}, we seek to understand how the underlying distribution of traits among competitors, and the relationship between traits and success influence the overall tournament. 
	
	This is an important question across disciplines. In biology the relationship between certain traits and success in competition for survival and reproduction is intrinsically related to fitness, and selection for heritable traits \cite{Stuart}. For example, competition for social dominance among male elephant seals depends on their body mass \cite{Haley} and competition among male dwarf Cape chameleons depends on coloration, head size, and body length \cite{Stuart}. Success in these competition events is correlated with reproductive success, suggesting that heritable traits which improve a male's chances of success are strongly selected for \cite{Haley}. In sports the relationship between the traits of a player or team and their success is an area of active interest - for athletes, owners, fans, and researchers alike. The rise of sabermetrics, the statistical study of baseball, is a popular example \cite{Lewis}. Sabermetrics have been used to predict the performance of players and teams based on their previous statistics. This includes the prediction of wins and losses as in \cite{Soto} where it was found that the success of a team depended on a variety of traits including batting average, fielding percentage, slugging percentage, and starting pitcher earned run average. 
	
	This paper answers questions 1 and 2 as follows:
	
	\begin{enumerate}

	\item[\textbf{1.}] 
	Rather than imposing the HHD framework ad hoc, we show that it arises naturally from the study of ranking and intransitivity.
	%The HHD is a natural method to choose since its use can be motivated without prior knowledge of the decomposition. 
	To illustrate this point, we provide a different derivation of the HHD than is provided by \cite{Jiang}. Instead of starting from the decomposition, we propose two special classes of tournaments with clear statistical motivation. 
	We then show that any tournament can be uniquely decomposed  into a combination of tournaments from these classes. This decomposition is the HHD (see Theorem \ref{thm: The HHD}). Next we illustrate that the HHD can be reached by six different approaches (Corollary \ref{Corollary: Equivalent formulations}), and is thus robust to varying motivations.

	\item[\textbf{2. }] We show that, under simple assumptions on the distribution of traits, the expected sizes of the components of the decomposition can be computed explicitly from the number of competitors, number of pairs who could compete, and the correlation in the performance of $A$ against $B$ with $A$ against $C$.  This correlation is shown to equal the uncertainty in the expected performance of a competitor. This relation links a decomposition of uncertainty in performance, to correlations in performance, and to tournament structure (see Theorem \ref{thm: expected transitivity and correlation coefficient} and Corollary \ref{Corollary: uncertainty in expectation}).
		
	\end{enumerate}
	
	The answers to the second question prove, under minimal assumptions, a series of intuitive statements about transitive/cyclic competition that appear, as heuristics, across the literature. These include:
	
	\begin{enumerate}
	\item
	\begin{enumerate}
	    \item The more predictable the performance of $A$ against a randomly drawn competitor (i.e., the less the performance of $A$ depends on their opponent) the more transitive the tournament.
	    \item 
	    \label{stm:1b}
	    The less predictable the performance of $A$ against a randomly drawn competitor (i.e., the more the performance of $A$ depends on their opponent) the more cyclic the tournament.
	 \end{enumerate}
	 \item
	 \begin{enumerate}
	    \item The more correlated the performance of $A$ against $B$ with the performance of $A$ against $C$, the more transitive the tournament.
	    \item The less correlated the performance of $A$ against $B$ with the performance of $A$ against $C$, the more cyclic the tournament.
	  \end{enumerate}
	   \item The more pairs of competitors who could compete, the more cyclic the tournament is, on average. 
	   \item Filling in missing data by random sampling overestimates intransitivity.
	\end{enumerate}

	The paper is structured as follows. In Section \ref{sec: Background} we provide some necessary background. Next, in Section \ref{sec: the Network HHD}, we derive the HHD in the context of tournaments and develop the associated ratings and intransitivity measure. In Section\ref{sec: trait-performance} we show how assumptions about the statistics underlying competition promote or suppress intransitivity. 
	We focus on trait-performance models in which performance is assumed to be a function of traits, which are sampled from a trait distribution. 
	We present a theorem (\ref{thm: expected transitivity and correlation coefficient}) which allows the expected size of the intransitivity measure to be computed directly from the number of competitors, edges in the network, and correlation in the performance of $A$ against $B$ with $A$ against $C$. This result is extended by a corollary (\ref{Corollary: uncertainty in expectation}) which shows that the correlation in performance is related to a decomposition in the uncertainty of the performance of $A$ against $B$. These results lead to a deeper conceptual understanding of how cyclic structure can arise from uncertainty in performance, and can be suppressed by correlation in performance. We present an example to illustrate the explanatory power of this theorem in Section \ref{sec: Case Studies}.
	
	\section{Background} \label{sec: Background}
	
	%% give necessary terminology
	Consider an ensemble of $m$ competitors. Assume that each competition event involves exactly two competitors, and never results in a tie. This standard assumption \cite{Kendall,Laird} can be weakened to allow for ties. We will refer to competition of this kind as a tournament.\footnote{This is distinct from a \textit{complete} tournament in which it must be possible for all pairs to compete.} 
	
	A tournament is specified by a schedule, and a set of win probabilities. The schedule fixes the order of events, and could be either fixed or random. For each possible pairing there is a pair of win probabilities. Let $p_{AB}$ denote the probability competitor $A$ beats $B$. The shorthand $A > B$ denotes the case when $A$ is expected to beat $B$ ($p_{AB} > 1/2$). It is the direction of competition. In principle the win probabilities could change in time, and could depend on the history of the process. We will focus on tournaments with unchanging win probabilities since evolving probabilities require additional modeling of temporal dynamics (see \cite{Glickman}). In addition we assume that the schedule and win probabilities are independent. We distinguish the structure of competition, which depends primarily on the win probabilities, from the dynamics of a tournament which depend on both the win probabilities and the schedule.
	
	The win probabilities may be conveniently represented using a competition network, $\mathcal{G}_{\rightleftarrows} = (\mathcal{V},\mathcal{E})$. Assign each competitor a node in the network. Introduce a pair of directed edges between each pair of competitors who could compete with each other. The edge from $B$ to $A$ is assigned the weight $p_{AB}$. In all that follows we will assume that the tournament is finite, \textit{connected} and \textit{reversible}. That is there are finitely many competitors, for any pair of competitors $A$ $B$ there is a path from $A$ to $B$ and from $B$ to $A$ through $\mathcal{G}_{\rightleftarrows}$ with probability greater than zero, and that  $p_{AB} \neq 0$ or $1$.
	
	Sometimes it is preferable to simplify the competition network by rounding all weights less than $1/2$ to $0$, and all weights greater than $1/2$ to $1$. This can be conveniently represented as an unweighted graph $\mathcal{G}_{\rightarrow}$ which contains all directed edges from $\mathcal{G}_{\rightleftarrows}$ with weights greater than a half, and an undirected edge between all pairs with $p_{AB} = 1/2$. This graph represents the expected direction of each competition event, as opposed to the probability of each event. Most intransitivity measures focus on this graph (see \cite{Kendall}, \cite{Landau_a}, \cite{Slater}).
	
	A \textit{ranking} is an ordered list of competitors from best to worst. This can be specified by a rank function $R$ which returns the rank of each competitor. Note that this is distinct from a \textit{rating}, $r$, which is a function that returns a real number for each competitor \cite{Langville}. Rankings are often generated by first generating a rating for each competitor, then listing them in decreasing order. Rankings and ratings provide an intuitive description of competition in which some innate competitive ability determines the performance of each competitor against all opponents.  
	
	Ranking methods are diverse, and well studied. Famous examples include the Page-rank method used by Google to sort search results \cite{Brin}, the Massey and Colley methods used by the NCAA to rank basketball and football teams \cite{Langville}, and the Elo rating/ranking widely used by chess federations \cite{Glickman,Stefani_c}. 
	The rating system produced by the HHD is a kind of log-least squares rating as is frequently used in paired comparison
	\cite{Bozoki,Kwiesielewicz_a,Kwiesielewicz_b}. 
	%\cite{Bozoki}, \cite{Kwiesielewicz_a}, \cite{Kwiesielewicz_b}. 
	Examples of least squares rating systems are included in \cite{Colley,Keener,Langville,Massey,Stefani_a,Stefani_b}.

	A competitive network $\mathcal{G}_{\rightleftarrows}$ is consistent with a ranking $R$ if $A > B$ whenever $R(A) < R(B)$. If a competitive network is consistent with a ranking then this ranking is unique and the network is \textit{transitive}. Transitive networks satisfy the intuitive property that if we consider some sequence of competitors with monotonically increasing rank, $A > B > C > D$ then $ A > D$. That is, $\mathcal{G}_{\rightarrow}$ contains no cycles, and all the edges in $\mathcal{G}_{\rightarrow}$ point from competitors who have high ranks (low ratings) to competitors with low ranks (high ratings). 
	
	If $\mathcal{G}_{\rightarrow}$ contains a cycle, then there exists a sequence of competitors such that $A > B > C > .... > A$, and the tournament is \textit{intransitive}. If a network is intransitive then it is not consistent with any ranking \cite{Petraitis}. Speaking broadly, measures of intransitivity either count the number of intransitive triangles present in $\mathcal{G}_{\rightarrow}$ \cite{Kendall}, or measure how far $\mathcal{G}_{\rightarrow}$ is from a nearby transitive network \cite{Slater}. The Kendall measure \cite{Kendall} counts the number of intransitive triangles in $\mathcal{G}_{\rightarrow}$. This can be done efficiently, however prioritizes triangles over larger loops and does not weight edges equally \cite{Appleby, Slater}. The Slater measure of intransitivity is the minimum number of edge directions that need to be reversed in order to transform $\mathcal{G}_{\rightarrow}$ into a transitive network \cite{Slater}. While conceptually preferable \cite{Jiang}, finding the closest transitive network is an NP hard problem  \cite{Bartholdi}, \cite{Endriss}, \cite{Hemaspaandra}, \cite{Jiang}. Despite some fast heuristics \cite{Eades}, this limits the application of the Slater measure to small networks. The intransitivity measure associated with the HHD is conceptually analogous to the Slater measure, but can be computed efficiently even for very large networks. Note that transitivity and intransitivity are defined relative to the direction of competition, that is, the \textit{sign} of $p_{AB} - 1/2$, rather than the exact value $p_{AB}$. In contrast the intransitivity measure associated with the HHD is continuous in the win probabilities, so uses all the information available in $\mathcal{G}_{\rightleftarrows}$.

	\section{The Network HHD} \label{sec: the Network HHD}
	
	The Network Helmholtz-Hodge Decomposition (HHD) can be derived by defining two special classes of tournaments.  These parallel the two classes of games defined in \cite{Candogan}.
	
	\subsection{Arbitrage Free and Favorite Free Tournaments}
	
	\subsubsection{Arbitrage Free Tournaments (Perfectly Transitive)}
	% The Cycle Condition: Arbitrage Free Tournaments
	
	A currency market is said to be \textit{arbitrage free} if it is impossible to make money by exchanging currencies in a cyclic fashion \cite{Jiang}. By analogy we define an \textit{arbitrage free tournament} to be a tournament for which it is impossible to expect to make money  by betting on cyclic sequences of events. Specifically, a tournament is arbitrage free if, for any cyclic sequence of competitors $\mathcal{C} = \{i_1,i_2,\ldots,.i_{n},i_{n+1} = i_{1} \}$, a sequence of wins where $i_{j}$ loses to $i_{j+1}$ ($i_1$ loses to $i_2$ loses to $i_3$ and so on) is equally likely as a sequence of wins where $i_{j}$ beats $i_{j+1}$ ($i_1$ beats $i_2$ who beats $i_3$ and so on). 
	This requires that the win probabilities satisfy a cycle condition.
	
		\textbf{Cycle Condition: } A tournament is arbitrage free if and only if, for every cycle $\mathcal{C} = \{i_1,i_2,\ldots,i_{n},i_{n+1} = i_1\}$, the win probabilities satisfy:
		\begin{equation} \label{eqn: Cycle Condition}
		p_{i_{1}i_{2}} p_{i_{2}i_{3}}...p_{i_{n}i_{1}} = p_{i_{1}i_{n}}...p_{i_{3}i_{2}}p_{i_{2}i_{1}}.
		\end{equation}
	
	The cycle condition can be expressed more simply by dividing the right hand side across to the left hand side and then taking a logarithm. This gives the equivalent condition:
	\begin{equation} \label{eqn: log Cycle Condition}
	\sum_{j = 1}^{n} f_{i_{j}i_{j+1}} = 0
	\end{equation}
	where the $f_{ij}$ is the log-odds that competitor $i$ beats competitor $j$:
	\begin{equation} \label{eqn: log odds edge flow}
	f_{ij} = \text{logit}(p_{ij}) = \log{\left(\frac{p_{ij}}{1 - p_{ij}} \right)}.
	\end{equation}
	
	Therefore the cycle condition is satisfied if and only if the sum of $f$ around any cycle is zero. The log-odds, $f$, are an example of an edge flow, an alternating function, $f_{ij} = -f_{ji}$, on the edges \cite{Jiang}. 
	
	% Properties of arbitrage free tournaments (elo and bradley terry satisfied exactly, bradley terry interpretable as steady state distribution, like page rank).
	
	\begin{snugshade}
	\textbf{Lemma 1: (Arbitrage Free) } \label{Lemma: Arbitrage Free Tournaments}
		A tournament is arbitrage free if and only if its win probabilities are consistent with a unique set of ratings $r$ that satisfy $p_{ij} = \text{logistic}(r_i - r_j)$ constrained to $\sum_{i} r_i = 0$ \footnote{$\text{logistic}(x) = \text{logit}^{-1}(x) = 1/(1 + \exp(-x))$.  }. Moreover if a tournament is arbitrage free then it is transitive.
	\end{snugshade}

	\textbf{Proof:}
	Suppose that a tournament is arbitrage free. Then it must satisfy the cycle condition. This implies that the sum of $f$ around any cycle is zero. It follows that, for any pair of endpoints $A, B$, the value of the sum of $f$ over a path connecting $A$ to $B$ is path independent.
	
\begin{figure}[t] \label{fig: spanning tree for potential}
		\centering
		\includegraphics[scale=0.35]{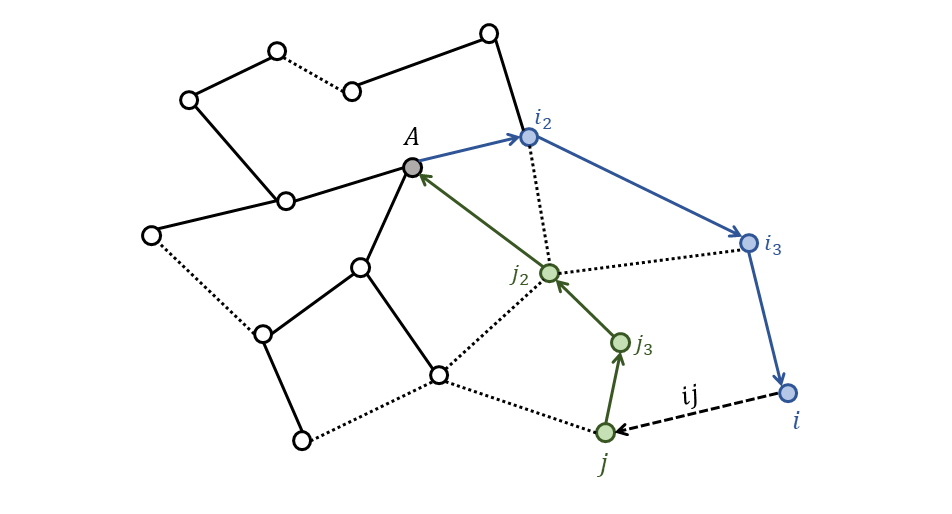}
		\caption{The spanning tree construction for recovering the ratings for an arbitrage-free tournament. The tree is shown with solid lines, and the chords with dotted lines. The root of the tree, $A$ is marked in grey. Two vertices, $i$ and $j$ connected by a chord $ij$, are shown in blue and green respectively. The sequence of nodes leading from $A$ to $i$ and $j$ are labelled. Then, by the cycle condition, the sum around the loop marked with arrows is zero, hence $f_{ij} = r_i - r_j$.}
	\end{figure}

	To recover the associated ratings, pick an arbitrary spanning tree of the network and an arbitrary starting competitor $A$.\footnote{A spanning tree is a subgraph of the network that contains no loops, includes all competitors, and is connected.}
	Then let $u_{B}$ equal the sum of $f$ over the path connecting $A$ to $B$ in the tree. Finally let $r_{B} = u_{B} - \frac{1}{m}\sum_{i} u_i$. 
	Then, by construction, $\sum_{i} r_i = 0$. It remains to show that $r_{i} - r_{j} = f_{ij}$ for all connected pairs $i,j$. 
	By construction, this must be true for all $i,j$ that are connected through an edge in the spanning tree. 
	Consider an edge not in the spanning tree (a chord) connecting $i$ and $j$. 
	Let $i_1 = A,i_2,\ldots,i_{l} = i$ and $j_1 = A,j_2,\ldots,j_{k}=j$ be the paths from $A$ to $i$ and $j$ through the spanning tree. Then $r_i - r_j = u_i - u_j = \sum_{n=1}^{l-1} f_{i_{n+1}i_{n}} - \sum_{n=1}^{k-1} f_{j_{n+1}j_{n}} = \sum_{n=k}^{2} f_{j_{n-1}j_{n}} + \sum_{n=1}^{l-1} f_{i_{n+1}i_{n}}$ which is the sum over the path from $j$ to $A$ then from $A$ to $i$. 
	If the chord was added to the path then this would complete a loop from $j$ to $A$ to $i$ back to $j$ (see Figure \ref{fig: spanning tree for potential}). By assumption the sum of $f$ around any loop is zero, so $r_i - r_j + f_{ji} = r_i - r_j - f_{ij} = 0$, or, $r_i - r_j = f_{ij}$. 
	Therefore, if a tournament is arbitrage free then there exist a set of ratings $r$ such that $r_i - r_j = f_{ij}$. 
	Since $f_{ij} = \text{logit}(p_{ij})$ this implies $p_{ij} = \text{logistic}(r_i - r_j)$. These ratings are unique since the sum of $f$ is path independent, hence the ratings generated by the spanning tree construction are independent of the choice of tree.
	
	Suppose that $p_{ij} = \text{logistic}(r_i - r_j)$. Then $f_{ij} = r_{i} - r_{j}$ for all connected $i,j$. This means that, given a path $i_1,i_2,\ldots,i_n$ the sum $f_{i_2 i_1} + f_{i_3 i_2} + ... f_{i_n i_{n-1}} = r_{i_n} - r_{i_1}$ as the sum is telescoping. If the path is a loop then $i_{n} = i_1$ so the sum equals zero. This means that $f$ satisfies the cycle condition, so the tournament is arbitrage free.
	
	Suppose the tournament is arbitrage free. Then $p_{ij} = \text{logistic}(r_i - r_j)$ for a unique set of ratings $r$. This means that $p_{ij} > 1/2$ if and only if $r_i > r_j$. It follows that $A > B$ if and only if $r_A > r_B$, so the win probabilities are consistent with the ranking induced by the ratings $r$. This means that the tournament is transitive. $\blacksquare$
	
    Lemma \ref{Lemma: Arbitrage Free Tournaments} shows that arbitrage free tournaments are the only tournaments which exactly match the logistic rating model $p_{ij} = \text{logistic}(r_i - r_j)$. This is the model assumed by the Elo rating system \cite{Aldous,Hvattum,Langville}.\footnote{The Elo rating system was originally proposed to rate chess players, but is also used to rank Sumo wrestlers \cite{Stefani_c}, English league football teams \cite{Hvattum} and international football teams. In the latter example the Elo method was the most predictive out of all methods tested \cite{Lasek}. The Women's World Cup uses a variant on the Elo method \cite{Lasek}.} 
	
	Arbitrage free tournaments are also the only tournaments which match the Bradley-Terry model:\footnote{The Bradley-Terry model is widely used in pairwise comparison and to rank competitors in tournaments. Examples include professional tennis \cite{McHale}, Cape dwarf chameleons \cite{Stuart} and northern elephant seals \cite{Haley}. Bradley-Terry models accounting for surface type, and discounting old games, have been shown to be effective in predicting the outcome of ATP tennis tournaments, consistently outperforming standard rankings \cite{McHale}. In a meta-study of predictive models the Bradley-Terry model had moderate predictive accuracy when compared to regression based methods, but was generally outperformed by Elo based methods which were the most accurate of all methods tested \cite{Kovalchik}.}  
	$p_{ij} = q_i/(q_i + q_j)$ where $q_i \geq 0$ for all $i$ \cite{Bradley_a,Bradley_b}. 
	If a network is arbitrage free, then from setting $q_i = \exp{(r_i)}$ it follows that  $p_{ij} = q_i/(q_i + q_j)$. 
	Alternatively, if the tournament satisfies the Bradley-Terry model, then setting $r_i = \log{(q_i)}$ produces a rating which satisfies $p_{ij} = \text{logistic}(r_i - r_j)$, so the network must be arbitrage free. 
	The values, $q$, which appear in the Bradley-Terry model are widely used as ratings. 
	
	Since arbitrage free networks are a special class of transitive networks, we will refer to these networks as ``perfectly" transitive. 
	Note that a perfectly transitive network must satisfy the cycle condition, which is a requirement on the values of $p$ rather than simply the sign of $p - 1/2$. 
	Hence, while all perfectly transitive networks are transitive, not all transitive networks are perfectly transitive. 
	For example, if $p_{AB} = 0.99$, $p_{BC} = 0.99$, and $p_{AC} = 0.51$ then the tournament is transitive, even though $p_{AC}$ is much smaller than might be expected given $p_{AB}$ and $p_{BC}$. 
	This tournament is not perfectly transitive since it does not satisfy the cycle condition. 
	
	% The Neighborhood Condition: favorite free Tournaments
	\subsubsection{Favorite Free Tournaments (Perfectly Cyclic)}
	
	In contrast to arbitrage free tournaments, we define a \textit{favorite free tournament} to be a tournament for which it is impossible to make money on average by betting on a favorite competitor over his or her neighbors. 
	Specifically, we require that in a favorite free tournament $A$ is equally likely to beat all of their neighbors, as to lose to all of their neighbors. This leads to a neighborhood condition.
	
		\textbf{ Neighborhood Condition: } A tournament is favorite free if and only if, for every competitor $i$ with neighborhood $\mathcal{N}(i)$, the win probabilities satisfy:
		\begin{equation} \label{eqn: neighborhood condition}
		\prod_{j \in \mathcal{N}(i)} p_{i j} = \prod_{j \in \mathcal{N}(i)} p_{j i}.
		\end{equation}
	
	Like the cycle condition, the neighborhood condition can be written directly as a condition on the log-odds edge flow $f$. 
	Dividing across by the left hand side and taking a logarithm we see that a tournament satisfies the neighborhood condition if and only if the sum of $f_{ij}$ over the neighborhood of $i$ is zero for all competitors $i$:
	\begin{equation} \label{eqn: neighborhood condition on f}
	\sum_{j \in \mathcal{N}(i)} f_{ij} = 0.
	\end{equation}
	
	If the neighborhood condition is satisfied then it can be extended to all sets of competitors. 
	Let $S$ be a set of competitors and let $\mathcal{N}(S)$ be the set of all competitors not in $S$ who neighbor $S$. 
	Then the neighborhood condition implies:
	\begin{equation} \label{eqn: neighborhood condition for sets}
	\sum_{j \in \mathcal{N}(S), i \in S} f_{ij} = 0.
	\end{equation}
	This identity follows from the discrete divergence theorem, which states that the sum of $f$ over the neighborhood of $S$ equals the sum of the divergence of every competitor in $S$. If $i$ and $j$ are both in $S$ then the sum over the neighborhood of $i$ contributes $f_{ij}$, and the sum over the neighborhood of $j$ contributes $f_{ji} = - f_{ij}$. 
	Therefore all the internal edges cancel in the sum. So $\sum_{j \in \mathcal{N}(S), i \in S} f_{ij} = \sum_{i \in S} \sum_{j \in \mathcal{N}(i)} f_{ij} = \sum_{i \in S} 0 = 0.$
	
	The cycle condition defined a special subset of transitive tournaments. The neighborhood condition also defines a special class that can be seen as a subset of a larger class - the class of \textit{cylic} tournaments.
	
	We define a \emph{cyclic tournament} to be a tournament such that, if there is a path from $A$ to $B$ in $\mathcal{G}_{\rightarrow}$, then there must be a path back from $B$ to $A$ in $\mathcal{G}_{\rightarrow}$. 
	
	\begin{snugshade}
	\textbf{Lemma 2: (Favorite Free)} \label{Lemma: favorite free Tournaments}
		A favorite free tournament is cyclic, and is never transitive unless $p_{ij} = 1/2$ for all connected $ij$.
	\end{snugshade}
	
	\textbf{Proof:}
	Suppose that a given tournament is favorite free. 
	Then $\sum_{j \in \mathcal{N}_i} f_{ij} = 0$ for all $i$. 
	This leaves two distinct possibilities, either $f_{ij} = 0$ for all $j \in \mathcal{N}(i)$, or there is some $j$ such that $f_{ij} \neq 0$. 
	The former case requires $p_{ij} = 1/2$ for all $j \in \mathcal{N}(i)$. 
	We will refer to this case as the \textit{neutral} case. 
	If the neighborhood of $i$ is not neutral then $f_{ij} \neq 0$ for some $j \in \mathcal{N}(i)$. 
	Since the sum over all $j$ is zero this means that there must be at least one other edge $ik$ such that $\text{sign}(f_{ij}) = - \text{sign}(f_{ik})$. 
	This means that, if there is an edge into competitor $i$ in $\mathcal{G}_{\rightarrow}$ there must also be at least one edge out of $i$ in $\mathcal{G}_{\rightarrow}$ (recall that if $p_{ij} = 1/2$ then there are a pair of edges between $i$ and $j$, one from $i$ to $j$ and one from $j$ to $i$). 
	
	Since the neighborhood condition can be extended from the neighborhood of competitors to the neighborhood of sets this property can also be extended to sets. That is, if there is an edge into the set $S$ in $\mathcal{G}_{\rightarrow}$ then there must also be an edge out of the set $S$ in $\mathcal{G}_{\rightarrow}$. 
	
	Now suppose that there is a path from $A$ to $B$ in $\mathcal{G}_{\rightarrow}$. It remains to construct a path back to $A$.
	
	\begin{figure}[t] \label{fig: favorite free implies cyclic}
		\centering
		\includegraphics[scale=0.35]{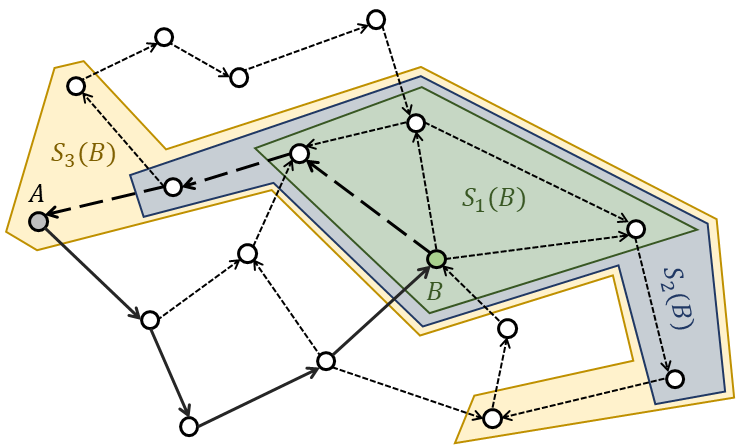}
		\caption{A favorite free tournament must be a cyclic tournament. 
		The arrows represent the direction of competition. 
		If the network is favorite free then if there is an edge pointing into a set there must be an edge pointing out of it. 
		A path from $A$ to $B$ is shown in black.  Then the sets $S_1(B), S_2(B), S_3(B)$ are shown as shaded polygons. These contain all competitors distance 1, 2, and 3 (respectively) from $B$. These sets continue to expand until they include $A$, hence there is a path from $B$ to $A$. }
	\end{figure}
	
	Define the nested sets $S_0(B), S_1(B),\ldots,$, where $S_{d}(B)$ is the set of all nodes that can be reached from $B$ with a path in $\mathcal{G}_{\rightarrow}$ of length less than or equal to $d$. 
	Now since there is a path from $A$ to $B$ in $\mathcal{G}_{\rightarrow}$ there is an edge in $\mathcal{G}_{\rightarrow}$ arriving at $\{B\} = S_{0}(B)$.
	Thus there is a path from $A$ to all competitors in $S_1(B)$. 
	Now there are two possibilities, either $A$ is in $S_1(B)$, or $A$ is not in $S_1(B)$. 
	If $A$ is in $S_1(B)$ then we are done. 
	If not, then there is an edge entering $S_1(B)$ in $\mathcal{G}_{\rightarrow}$ since there is a path from $A \notin S_1(B)$ to $B \in S_1(B)$. 
	Then the neighborhood condition implies that there is an edge out of $S_1(B)$, which means that $S_2(B) \neq S_1(B)$. 
	Now the logic repeats. Either $A$ is in $S_2(B)$, in which case we are done, or it is not.
	If it is not then there must be an edge entering $S_2(B)$ so there must be an edge leaving $S_2(B)$ so $S_3(B) \neq S_2(B)$. 
	This means that, as long as $A \notin S_{d}(B)$ there is a larger set $S_{d+1}(B) \neq S_{d}(B)$ which can be reached from $B$.  Since we assumed that there are finitely many competitors this can only continue until $A$ is contained in $S_{d}(B)$ for some $B$. This proof technique is illustrated in Figure \ref{fig: favorite free implies cyclic}.
	
	Suppose that the tournament is transitive, favorite free, and not neutral. Since it isn't neutral there must be at least one pair $ij$ such that $p_{ij} > 1/2$. This means that $r_i > r_j$ and there is an edge from $j$ to $i$ in $\mathcal{G}_{\rightarrow}$. But, if the tournament is favorite free then there must be some other path from $i$ back to $j$ in $\mathcal{G}_{\rightarrow}$. This means that $r_j > r_i$ since there is a path in $\mathcal{G}_{\rightarrow}$ from $j$ to $i$. This is clearly a contradiction. This implies that a cyclic tournament is not transitive unless it is neutral: $p_{ij} = 1/2$ for all $ij$.\footnote{This shows that the two classes of tournaments are distinct, as their only overlap is the neutral case. Note that a neutral tournament is considered transitive since it can be consistently ranked - all competitors should be ranked the same. }
	$\blacksquare$
	
	So, just as the cycle condition (no tendency to cycle) implied transitivity, the neighborhood condition, (no favorites) implies that the network is cyclic, and is only transitive if it is also completely neutral. As before, whether a tournament is cyclic or not depends on the sign of $p_{ij} - 1/2$, while the neighborhood condition is a condition on the values of $p_{ij}$. This motivates the definition: a tournament is \textit{perfectly cyclic} if and only if it is favorite free. As before, all perfectly cyclic tournaments are cyclic, but not all cyclic tournaments are perfectly cyclic.
	
	Note that, unlike perfectly transitive tournaments where $f$ is determined by a set of ratings $r$, we are not currently equipped to relate the edge flow of a favorite free tournament to a lower dimensional representation. In Section \ref{sec: the HHD} we will show that a favorite free tournament has edge flows $f$ which can always be represented as a sum of cyclic intensities (or vorticities) on a set of loops. This result will parallel the conclusions of Lemma \ref{Lemma: Arbitrage Free Tournaments}. 
	
	\subsection{The Discrete HHD}
	
	Given these two classes of tournaments it is natural to ask: can a generic tournament be decomposed into a perfectly transitive (arbitrage free) part and a perfectly cyclic (favorite free) part? We answer in the affirmative. This is the Helmholtz-Hodge decomposition.
	
	\subsubsection{Operators}
	% Gradient, and Curl
	
	In order to define the decomposition succinctly it is helpful to have a pair of operators analogous to the gradient and curl operators in the continuum. 
	We simplify the topological presentation in \cite{Jiang} by expressing the decomposition entirely through linear algebra. 
	
	First, we define the edge space $\mathbb{R}^{E}$, where $E$ is the number of pairs $i,j$ who could compete. 
	Index each pair so that edge $k$ has endpoints (competitors) $i(k),j(k)$. 
	Note that this requires assigning each edge an arbitrary start and endpoint so that positive $f$ indicates motion from the start to the end, while negative $f$ indicates motion from the end to the start. 
	This is simply a sign convention.
	
	Let the \textit{discrete gradient} operator $G$ be the matrix which maps from $\mathbb{R}^m$ to $\mathbb{R}^E$ by setting:
	\begin{equation} \label{eqn: discrete gradient}
	[G u]_{k} = u_{i(k)} - u_{j(k)}.
	\end{equation}
	
	Notice that if $r$ is a rating function on the nodes, then attempting to find $r$ such that $r_i - r_j = f_{ij}$ is equivalent to looking for $r$ such that $Gr = f$. 
	Since any arbitrage free tournament admits a unique rating $r$ such that $G r = f$ it follows that the space of perfectly transitive networks is equivalent to the space of tournaments with edge flow $f$ in the range of the gradient. Assuming that the tournament is connected, the gradient has a one-dimensional null-space parallel to the vector $[1;1;...1]$. It follows that $G(r + c) = Gr$ if $c$ is some constant. This motivates the constraint $\sum_i r_i = 0$ used throughout, since the edge flow only determines the size of differences in ratings, not the actual ratings. 
	
	The gradient transpose, $G^T$ is the discrete divergence operator. The divergence maps from the space of edges to the space of nodes (competitors) such that:
	\begin{equation} \label{eqn: discrete divergence}
	[G^T f]_i = \sum_{\mathcal{N}(i)} f_{ij}.
	\end{equation}
	
	The neighborhood condition, \ref{eqn: neighborhood condition on f}, is equivalent to requiring that $G^T f = 0$. That is, the space of favorite free tournaments is equivalent to the space of tournaments with edge flow $f$ in the null space of the divergence. Note that, like the divergence operator in the continuum, the discrete divergence obeys the divergence theorem (the sum of the divergence on the neighborhood of each competitor in a set is the same as the sum of the edge flow into the set).
	
	\begin{figure}[t] \label{fig: operators}
		\centering
		\includegraphics[scale=0.3]{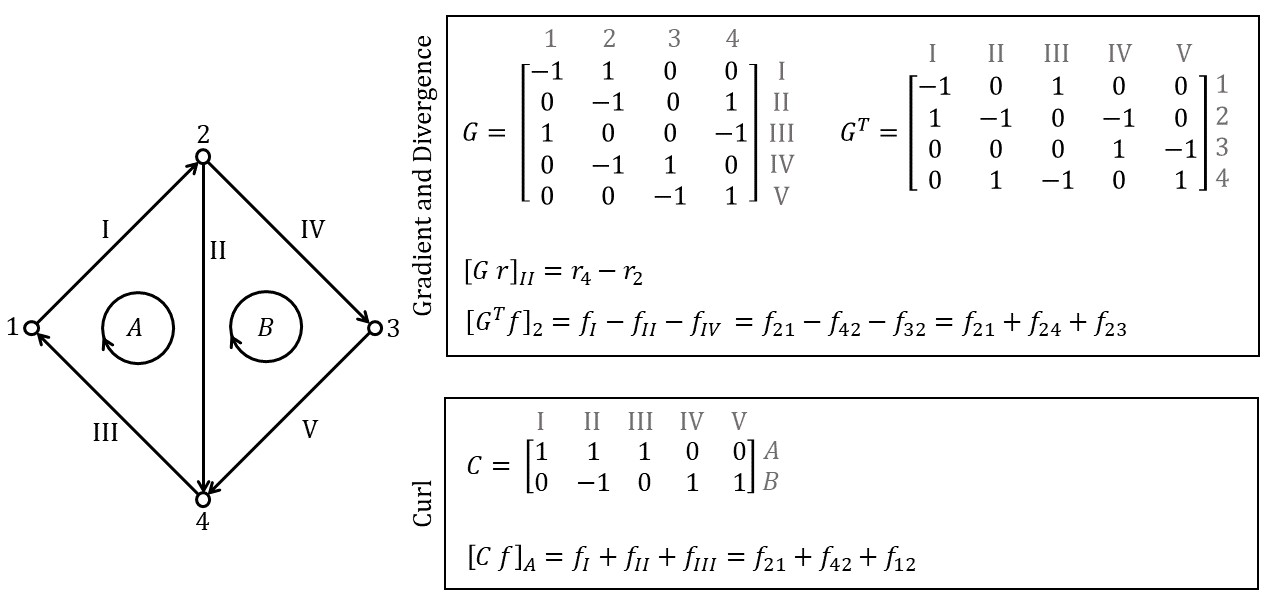}
		\caption{The gradient, divergence, and curl for an example network. }
	\end{figure}
	
	In order to build a parallel description for perfectly cyclic tournaments, we need a space of loops. First define the sum of two cycles $\mathcal{C}_1$, $\mathcal{C}_2$ to be all edges included in either $\mathcal{C}_1$ or $\mathcal{C}_2$ but not both. Equipped with this addition operation, the space of cycles is a vector space, which can be represented with a cycle basis. A \textit{cycle basis} is a collection of linearly independent cycles $\mathcal{C}_1, \mathcal{C}_2,\ldots,\mathcal{C}_L$ such that any other cycle $\mathcal{C}$ can be expressed as a linear combination of cycles in the loop basis \cite{Galbiati}.  
	
	Any connected graph admits a cycle basis. A constructive method for finding a cycle basis follows. First, pick a spanning tree of the network. Then the spanning tree includes $m-1$ edges, and $E - (m-1)$ edges are left out. These are the \textit{chords}. 
	By construction, the tree does not contain any loops. If one chord is added to the tree then the network contains exactly one cycle. Note that no two chords can produce the same cycle, and that the set of cycles produced by adding the chords to the spanning tree is necessarily linearly independent since no chord appears in more than two of these cycles. 
	Therefore, if we enumerate the chords from $1,2,\ldots,L = E - m + 1$ then the set of cycles $\mathcal{C}_1,\ldots,\mathcal{C}_L$ associated with each chord is a cycle basis. A basis generated by a spanning tree is a \textit{fundamental cycle basis} \cite{Bollobas, Galbiati}. 
	This basis is not unique, since there are often many different possible spanning trees, moreover not all cycle bases need be constructed via a spanning tree.\footnote{A collection of cycles $\mathcal{C}_1,\ldots,\mathcal{C}_n$ with $n > L$ that spans the space of loops is an overdetermined cycle basis. In practice we can work with either a cycle basis or an overdetermined cycle basis depending on the application \cite{Jiang}. This flexibility can be useful. For example, given a complete graph the set of all triangles that include a particular competitor forms a cycle basis, while the set of all triangles forms an overdetermined basis. If there is no reason a priori to identify a particular competitor as special then it may be more natural to work in the overdetermined basis of all triangles.}
	
	Next define the cycle space $\mathbb{R}^{L}$ to be the space of real vectors with one entry for each cycle in a chosen cycle basis. 
	The dimension of the cycle space $L = E - m + 1$ is the \textit{cyclomatic number} of the network \cite{Bollobas, Galbiati}. 
	Then we define the \textit{discrete curl} operator to be the matrix which maps from $\mathbb{R}^E$ to $\mathbb{R}^L$ (edges to cycles) by summing $f$ around each loop. That is, if the set of edges $\{k_1,k_2,\ldots,k_{n_l}\} = \mathcal{C}_{l}$ then:
	\begin{equation} \label{eqn: Curl operator}
	[C f]_{l} = \sum_{h=1}^{n_l} f_{i(k_{h}) j(k_{h})}.
	\end{equation}
	
	Note that in order to perform this sum, each loop must be assigned an arbitrary direction of traversal. This is simply a sign convention.
	
	In general we will only consider curl operators that are defined with respect to cycle bases such that there exists an invertible $L\times L$ matrix $T$ for which $TC = \tilde{C}$, where $\tilde{C}$ is the curl operator defined with respect to a fundamental cycle basis. 
	
	The curl is analogous to the curl in continuous space, which is a path integral over infinitesimally small loops. Note that the discrete curl defined in this way is more general than the discrete curl defined in \cite{Candogan} or \cite{Jiang}. 
	Jiang and Candogan restrict the curl operator to only act on connected cliques of three nodes (triangles), and then introduce additional operators to account for cliques containing more nodes. 
	This construction can lead to unintuitive conclusions. For example, if $p_{AB} = p_{BC} = p_{CD} = p_{DA} = 0.99$ then there is clearly a cyclic tendency in the competition, but if the curl is restricted to only act on triangles, then the curl of this graph is zero. Here we extend the curl to act on loops of arbitrary length since, like \cite{Slater}, we do not see a fundamental distinction between cyclic structure on triangles and cyclic structure on larger loops. 
	If desired, we could partition the curl operator into blocks, each according to loops of a fixed length, and treat each block as the curl operator restricted to loops of a given size.
	
	The operators for an example network are provided in Figure \ref{fig: operators}.
	
	\begin{snugshade}
	\textbf{Lemma 3: (Orthogonality)} \label{Lemma: C and G are orthogonal}
		The curl $C$ and the gradient $G$ are orthogonal, regardless of the choice of cycle basis.
	\end{snugshade}
	
	\textbf{Proof:}
    Consider the product $C G u$ for some arbitrary vector $u \in \mathcal{R}^m$. The product $G u$ produces an edge flow, so the product $C G u$ produces a vector whose entries are the sum of that edge flow around a set of loops. Consider an arbitrary path $i_1,i_2,\ldots,i_n$. Then the sum of $G u$ over the path is $(u_{i_2} - u_{i_1}) + (u_{i_3} - u_{i_2}) + ... (u_{i_n} - u_{i_{n-1}}) = u_{i_n} - u_{i_1}$. Therefore, if the path is a loop, $i_n = i_1$ so the sum is zero. It follows that $C G u = 0$ for all $u \in \mathbb{R}^n$ so:
	\begin{equation} \label{eqn: orthogonality}
	CG =0, \quad G^T C^T = 0
	\end{equation}
	where the second equation follows trivially by transposing the first equation.\footnote{Note that the product $GC$ has no meaning in our framework. Even if the range of $C$ and domain of $G$ were of compatible dimension, the product has no natural interpretation since $C$ maps to loops and $G$ acts on nodes.} 
	$\blacksquare$
	
	\begin{snugshade}
	\textbf{Lemma 4: } \label{Lemma: nullspace of C subset of range of G}
		If $C$ is a discrete curl operator then if $Cf = 0$, there exists a set of ratings $r$ such that $Gr = f$.
	\end{snugshade}

	\textbf{Proof:} This is a direct consequence of Lemma \ref{Lemma: Arbitrage Free Tournaments}. If $C$ is a curl operator, then there exists an invertible transform $T$ such that $C = T \tilde{C}$ where $\tilde{C}$ is the curl operator with respect to some fundamental cycle basis. Then $C f = T \tilde{C} f = 0$ if and only if $\tilde{C} f = 0$. 
	Since $\tilde{C}$ is defined with respect to a fundamental cycle basis, $\tilde{C}$ is defined with respect to a spanning tree $\mathcal{T}$ which generates the cycle basis. Requiring that $\tilde{C} f = 0$ is equivalent to requiring that the sum of $f$ around every loop formed by adding one chord into the tree is zero. This condition is sufficient to reconstruct $r$ such that $Gr = f$ using the spanning tree construction given in the proof of Lemma \ref{Lemma: Arbitrage Free Tournaments}, where the chosen tree is $\mathcal{T}$. $\blacksquare$
	
	Lemma \ref{Lemma: C and G are orthogonal} and Lemma \ref{Lemma:  nullspace of C subset of range of G} establish that any $f$ in the range of the gradient is in the nullspace of the curl, and any $f$ in the nullspace of $C$ is in the range of the gradient. That is, if $f = Gr$ then $Cf = 0$ and if $Cf = 0$ then $f = Gr$ for some rating $r$. 
	Therefore the range of the gradient is the nullspace of the curl. The equivalence of these two spaces and the orthogonality of the operators allows us to decompose $f$ into unique perfectly transitive and perfectly cyclic components. This is the HHD. 
	
	\subsubsection{The Discrete Helmholtz-Hodge Decomposition} \label{sec: the HHD}
	
	We are now equipped to prove that every edge flow can be represented as the sum of a perfectly transitive (arbitrage free), and perfectly cyclic (favorite free) edge flow - thus any tournament can be represented as a unique combination of a perfectly transitive and perfectly cyclic tournament. Similar proofs are provided in \cite{Candogan} and \cite{Jiang} .

	\begin{snugshade}
	\textbf{Theorem 5: (The HHD)} \label{thm: The HHD}
		 Any $f \in \mathbb{R}^{E}$ can be decomposed such that:
		\begin{equation} \label{eqn: decompose f}
		f = f_t + f_c
		\end{equation}
		where $f_t$ is arbitrage free (perfectly transitive) and $f_c$ is favorite free (perfectly cyclic):
		\begin{equation} \label{eqn: nullspaces}
		C f_t = 0, \quad G^T f_c = 0.
		\end{equation}
		and both are unique. In addition, there exists a unique rating $r$ satisfying $\sum_i r_i = 0$ such that $f_t = G r$
		and for any choice of cycle basis there exists a unique vorticity $v \in \mathbb{R}^L$ such that $f_c = C^T v.$
        Thus the original edge flow $f$ can be uniquely decomposed:
		\begin{equation} \label{eqn: HHD}
		f = G r + C^T v.
		\end{equation}
		
	\end{snugshade}
	
	\textbf{Proof: } 
	By the fundamental theorem of linear algebra (Fredholm alternative):
	\begin{equation} \label{eqn: Fredholm}
	\mathbb{R}^E = \text{null}(G^T) \oplus \text{range}(G).
	\end{equation}
	
    Lemma \ref{Lemma: C and G are orthogonal} and Lemma \ref{Lemma: nullspace of C subset of range of G} guarantee that $\text{range}(G) = \text{null}(C)$, so:
	\begin{equation} \label{eqn: arbitrage free plus favorite free}
	\mathbb{R}^E = \text{null}(G^T) \oplus \text{null}(C).
	\end{equation}
	
	This establishes equation \ref{eqn: decompose f}, where $f_t$ is the orthogonal projection of $f$ onto $\text{null}(C)$ and $f_c$ is the orthogonal projection of $f$ onto $\text{null}(G^T)$.
	
	To prove that the arbitrage free and favorite free fields can be expressed using ratings and vorticities, write:
	\begin{equation}
	\mathbb{R}^{E} = \text{null}(C) \oplus \text{range}(C^T).
	\end{equation}
	
	Then using $\text{null}(C) = \text{range}(G)$:
	\begin{equation} \label{eqn: edge space is perfectly transitive plus perfectly cyclic}
	\mathbb{R}^E = \text{range}(G) \oplus \text{range}(C^T).
	\end{equation}
	
	Equation \ref{eqn: edge space is perfectly transitive plus perfectly cyclic} means that there exists an $r$ such that $G r = f_t$, and there exists a $v$ such that $C^T v = f_c$. We have already proved $r$ was unique. 
	To prove that $v$ is unique we use rank nullity. Equation \ref{eqn: edge space is perfectly transitive plus perfectly cyclic} guarantees $E = \text{rank}(G)+\text{rank}(C^T)$. 
	In general $G$ has rank $m - 1$ since the Laplacian, $G^T G$, has nullity equal to the number of connected components in the network \cite{Bollobas}. We assumed the network is connected, so $G^T G$ has nullity 1, thus $G$ has a one-dimensional nullspace. This nullspace corresponds to the vector of all ones, since the gradient of a constant is zero. Therefore $\text{rank}(C^T) = E - (m - 1) = L$. By construction, $C^T$ has $L$ columns, therefore $C^T$ is full rank. It follows that the linear system  $C^T v = f$ has a unique solution if $f \in \text{range}(C^T)$.\footnote{This result could also be obtained more intuitively as follows. Note that if $C$ is defined with respect to a fundamental cycle basis, then by ordering the edges so that all of the chords are indexed before all of the edges in the tree, the operator $C$ is a block matrix whose first $L \times L$ block is the identity. It follows that $C$ is in row reduced echelon form and has rank $L$. The column rank of a matrix is its row rank so the rank of $C^T$ is also $L$.} $\blacksquare$
	
	\begin{figure}[t] \label{fig: HHD diagram}
		\centering
		\includegraphics[scale=0.3]{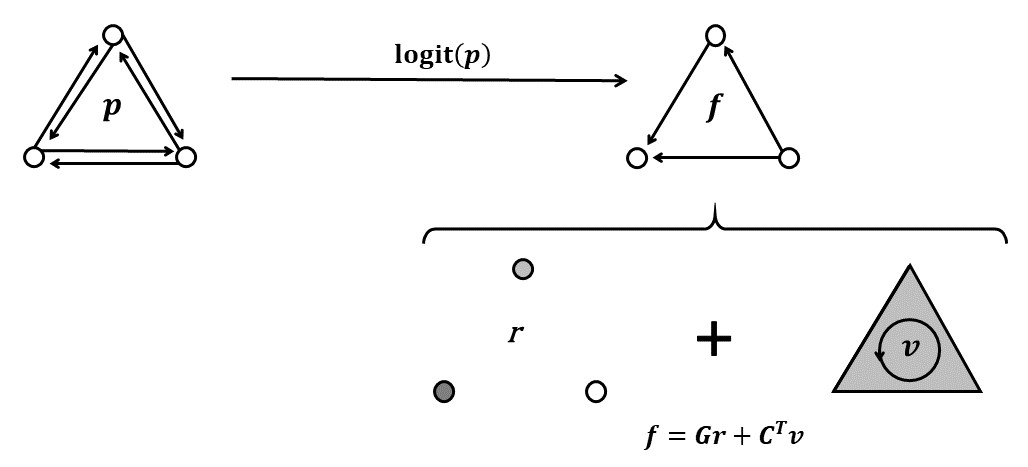}
		\caption{A schematic representation of the decomposition for a complete tournament on three competitors. The edge flow $f$ is set equal to $\text{logit}(p)$, and then broken into a set of ratings $r$ and vorticities $v$, such that $f = G r + C^T v$.}
	\end{figure}
	
	This proves that an arbitrary tournament can be decomposed into a perfectly transitive and a perfectly cyclic tournament, where the perfectly transitive tournament is specified by a set of ratings, and the perfectly cyclic tournament is specified by a set of vorticities. The ratings associated with the HHD are the Hodge ratings proposed by \cite{Jiang}. Figure \ref{fig: HHD diagram} provides a schematic representing the decomposition.
	
	The gradient $G$ has exactly 2 nonzero entries per edge, so it becomes sparser as the number of competitors increases.  Consequently, the decomposition can be performed efficiently, even for large, fully connected networks. Methods are discussed in \cite{Candogan,Jiang}. 
	
	The intransitivity measure associated with the HHD is the size of the cyclic component $||f_c||_2$. Because the HHD is a decomposition onto orthogonal subspaces, this measure is equal to the distance from $f$ to the closest perfectly transitive tournament. Therefore the Helmholtz-Hodge intransitivity measure is conceptually analogous to the Slater intransitivity measure \cite{Slater}, and its variants \cite{Petraitis}, \cite{Soliveres}, \cite{Ulrich}. Similarly, the transitivity measure associated with the HHD is the size of the transitive component $||f_t||_2$, and is the distance from $f$ to the closest perfectly cyclic tournament.
	
	\begin{figure}[t] \label{fig: Transitive Intransitive Plane}
		\centering
		\includegraphics[scale=0.2]{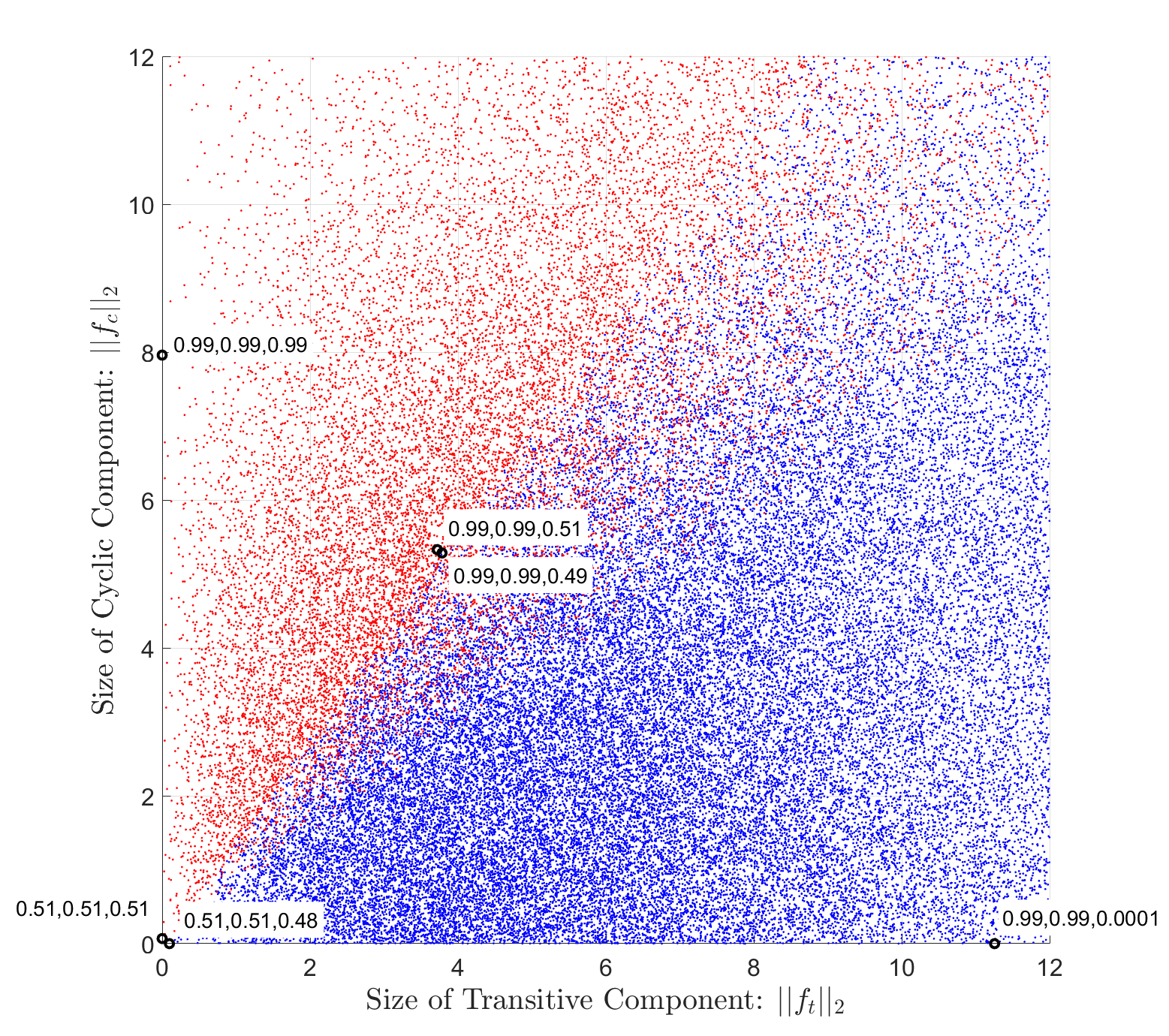}
		\caption{ Transitivity and intransitivity of $10^4$ triangular networks with randomly drawn win probabilities. The horizontal axis is the size of the transitive component and the vertical axis is the size of the cyclic component. Each scatter point is a sampled network. Blue scatter points are transitive, red are intransitive. The large black circles represent example networks. The text next to each example gives the probability $A$ beats $B$, $B$ beats $C$, and $C$ beats $A$. If all of these numbers are greater than $0.5$ then the network is intransitive. Note that the classification into transitive and intransitive draws a sharp distinction between networks whose win probabilities are nearly identical, while networks with similar win probabilities remain close to each other when using the Hodge measures. Also note that the boundary between transitive and intransitive networks is an angular sector, hence this classification is based on the relative sizes of the transitive and cyclic components, not their absolute sizes. In contrast the Hodge measures reflect the absolute size of each component. Thus the example with win probabilities $0.99,0.99,0.49$ can be transitive and the example $0.51,0.51,0.51$ can be intransitive, even though the former has a larger cyclic component than the latter.}
	\end{figure}

	Note that these measures are continuous in $p$. This sets the measure associated with the HHD apart from classical methods which depend only on the direction of competition encoded in $\mathcal{G}_{\rightarrow}$ such as the Kendall \cite{Kendall} or Slater \cite{Slater} measures. These methods are discrete in $p$. 
	This distinction is important, since it means that the Helmholtz-Hodge measure distinguishes between the cases $p_{AB} = p_{BC} = p_{CA} = 0.99$ and $p_{AB} = p_{BC} = p_{CA} = 0.51$ (intransitivity $7.96$ and $0.07$ respectively). Using the discrete measures, these two tournaments are equally intransitive. Thus the Helmholtz-Hodge measure is distinguishes between strong and weak intransitive cycles, and so reflects the absolute strength of cyclic competition. The discrete measures reflect the relative strength of cyclic competition since they only depend on the sign of $f$, which depends on both $f_c$ and $f_t$. If the transitive part is large then it may mask weaker cyclic competition when using a discrete measure. 
	For example, if $p_{AB} = 0.99, p_{BC} = 0.99$ and $p_{CA} = 0.49$ then it is clear that the probability that $C$ beats $A$ is much larger than might be expected using any predictive rating of the competitors. 
	However, in this example competition is transitive so all discrete measures of intransitivity would return their minimal value, 0. In contrast, the Helmholtz-Hodge measure returns intransitivity $5.29$. These examples are illustrated in Figure \ref{fig: Transitive Intransitive Plane}
	Normalizing the Helmholtz-Hodge measures by $||f||_2$ produces the equivalent relative measures: $||f_c||_2/||f||_2$ and $||f_t||_2/||f||_2$. 
	
	\subsubsection{Equivalent Formulations} \label{sec: Equivalent Formulations}
	
	Here we present six different approaches that arrive at the same decomposition. These provide different and useful perspectives on the HHD, and illustrate that it is robust to varying motivations. The ensuing Corollary follows directly from standard properties of projection onto orthogonal subspaces, so we omit the proof.
	
	\begin{snugshade}
	\textbf{Corollary 6: (Equivalent Formulations)}
	\label{Corollary: Equivalent formulations}
		The following six decompositions are equivalent:
		\begin{enumerate}
			\item $f = f_t + f_c$ where $f_t$ is arbitrage free and $f_c$ is favorite free; 
			\item $f = f_t + f_c$ where $f_t = G r$ for some rating $r$ and $f_c = C^T v$ for some vorticity $v$;
			\item the ratings $r$ satisfy:
			\begin{equation} \label{eqn: least squares for HHD ratings}
			r = {\rm{argmin}}_{u|\sum_i u_i = 0}\left\{|| G u - f||^2_2\right\} 
			\end{equation}
			and set $f_t = G r, f_c = f - f_t$;
			
			\item the vorticities $v$ satisfy:
			\begin{equation} \label{eqn: least squares for HHD voriticies}
			v = {\rm{argmin}}_{v}\{|| C^T v  - f||^2_2\}
			\end{equation}
			and set $f_c = C^T v, f_t = f - f_c$;
			
			\item $f = f_t + f_c$ where $f_t = G r$ for the unique ratings $r$ such that the circulant $f - f_t$ is favorite free;
			
			\item $f = f_t + f_c$ where $f_c = C^T v$ for the unique vorticities $v$ such that $f - f_c$ is arbitrage free.
			
		\end{enumerate}
	\end{snugshade}
	
    The first decomposition separates $f$ into a pair of flows each defined by what it is not: namely, one is not circulatory, and the other has no tendency to diverge or converge. The second decomposition separates $f$ into a pair of flows each defined by what they are: namely, one is perfectly transitive, and the other is perfectly cyclic. The equivalence of these two decompositions was established by Theorem \ref{thm: The HHD}. 
	
	The next two decompositions are based on fitting problems. In each case the goal is to represent $f$ as nearly as possible when restricted to the range of an operator. Decomposition 3 searches for a set of ratings $r$ such the error, $Gr - f$, is minimized in the least squares sense. This means that the ratings produced by the HHD are a type of least squares rating, in particular, log least squares rating \cite{Bozoki, Kwiesielewicz_a,Kwiesielewicz_b}. Least squares ratings methods are widely used \cite{Colley,Keener,Langville,Massey,Stefani_a,Stefani_b}.
	Decomposition 3 also shows that the HHD is equivalent to finding the nearest perfectly transitive edge flow. 
	
	Similarly, Decomposition 4 searches for a set of vorticities $v$ such that the error $C^T v - f$ in approximating $f$ with $C^T v$ is minimized in the least squares sense. This is equivalent to finding the nearest perfectly cyclic edge flow.
	Although the literature has focused almost exclusively on Decomposition 3, decompositions 3 and 4 are dual to one another. This parity in approach sets the HHD apart from existing methods.
	
	The final two decompositions are defined by enforcing a constraint on the residue when approximating $f$ with either the gradient of a set of ratings or the curl transpose of a set of vorticities. These approaches can be motivated as follows. Suppose one sought a rating $r$ such that $Gr$ approximated $f$. 
	The error in this approximation (the circulant) is $Gr - f$. 
	As long as the divergence of the circulant is nonzero the approximation has not captured a tendency of the edge flow to either point inwards towards, or outwards from, a competitor. If the net flow into a competitor is positive, then that competitor tends to outperform their neighbors in a way that the ratings fail to capture. Therefore it would be natural to adjust the ratings until the net flow into or out of any set of competitors is zero. That is, until the divergence of the circulant is zero, or equivalently, the circulant is favorite free. 
	
	The final decomposition can be motivated similarly. Define the \emph{divergent}, $C^T v - f$ to be the error upon approximating $f$ with vorticity $v$. 
	As long as the curl of the divergent is nonzero, the approximation has failed to capture some tendency of $f$ to circulate. This tendency to circulate is exactly what the vorticities are meant to capture, so it is natural to look for a $v$ such that the curl of the divergent is zero on every loop. That is, until the divergent is arbitrage free.
	
	Corollary \ref{Corollary: Equivalent formulations} shows that the decomposition into arbitrage free and favorite free components, perfectly transitive and perfectly cyclic components, the nearest perfectly transitive approximation, the nearest perfectly cyclic approximation, the perfectly transitive approximation with favorite free circulant/error, or the perfectly cyclic approximation with arbitrage free divergent/error, are all the same. The fact that the HHD is equivalent to all of these different approaches motivates its use.

	\section{The Trait-Performance Theorem} \label{sec: trait-performance} 
	
	How intransitive is a typical tournament? Using the intransitivity measure associated with the HHD, this question is the same as asking, how cyclic is a tournament on average?
	
	Answering this question requires defining a statistical model for sampling tournaments - in particular, for sampling edge flows. How do assumptions about the distribution of possible edge flows affect the expected strength of cyclic competition? What statistical features tend to promote or suppress cyclic competition? 
	
	We initially explore these questions for a generic null model in which the edge flow, $F$, is sampled randomly from an unspecified distribution. This analysis identifies which statistical features of the edge flow, and which features of the network topology, influence the expected strength of cyclic competition. This sets the stage for our main result. If the edge flow is sampled using a trait-performance model, then the correlation structure of the edge flow takes on a canonical form which depends only on \textit{two} statistical quantities: the variance in the flow on each edge, and the correlation in the flow on pairs of edges that share an endpoint. This simplified correlation structure allows us to express the expected sizes of the cyclic and transitive components in a simple closed form that separates the influence of the network topology from the chosen trait-performance model.
	
	\subsection{Generic Null Models} \label{Generic Null Models}
	
	We start by considering a generic null-model for the edge flows $f$. Let $F \in \mathbb{R}^E$ be a random edge flow drawn from some distribution. Assume that the expected edge flow $\bar{f} = \mathbb{E}[F]$ is known, as is the covariance $V = \mathbb{E}[(F - \bar{f})(F - \bar{f})^T]$. 
	
	Let $P_c$ be the orthogonal projector onto the space of perfectly cyclic (favorite free) tournaments. Then the expected absolute strength of cyclic competition is:
	\begin{equation} \label{eqn: generic null model}
	\begin{aligned}
	\mathbb{E}[||F_c||^2] & = \mathbb{E}[F^T P_c^T P_c F] = \mathbb{E}[F^T P_c F] = \mathbb{E}\left[\sum_{kl} \left(P_c\right)_{kl} F_k F_l \right] = \\
	& \sum_{kl} \left(P_c\right)_{kl} \mathbb{E}[F_k F_l] = \sum_{kl} \left(P_c\right)_{kl} (\bar{f}_k \bar{f}_l + v_{kl}) = ||\bar{f}_c||^2 + \text{trace}(P_c V)
	\end{aligned}
	\end{equation}
	where $||\bar{f}_c||^2 = \bar{f}^T P_c \bar{f}$ and $\text{trace}(P_c \Sigma) = \sum_{kl} 
	\left(P_c\right)_{kl} v_{kl}$ is the matrix inner product between the projector and the covariance matrix.
	
	Therefore, no matter the underlying distribution of edge flows, the expected strength of cyclic competition is determined exclusively by three quantities: the \textit{expected edge flow}, the \textit{covariance in the edge flow}, and the \textit{topology of the network} (which determines $P_c$).
	
	The matrix inner product can be simplified if the flows on each edge are independent. Then $V$ is diagonal with entries $\sigma_{k}^2 = \mathbb{E}[(F_k - \bar{f}_k)^2]$. It follows that $\text{trace}(P_c V) = \sum_{k=1}^{E} \left(P_c\right)_{kk} \sigma_k^2$.
	
	The nonzero eigenvalues of a projector all equal one, so its trace equals the dimension of the space it projects onto. The projector $P_c$ projects onto the space of perfectly cyclic tournaments, which has dimension $L = E - (m-1)$. Therefore $\sum_k \left(P_c\right)_{kk} = L$.  Rewrite the expected strength of cyclic competition:
	\begin{equation} \label{eqn: independent edges null model cyclic}
	\mathbb{E}[||F_c||^2] = ||\bar{f}_c||^2 + L \sum_{k=1}^{E} \left(\frac{\left(P_c\right)_{kk}}{L} \right) \sigma_k^2.
	\end{equation}
	
	Since the diagonal entries of an orthogonal projector are always nonnegative, the right hand term can be interpreted as a weighted average of the variance on each edge. 
	Therefore, when the edges are independent, the expected strength of cyclic competition is given by the strength of the cyclic component of the expected edge flow, plus the dimension of the loop space times a weighted average of the variance on each edge. 
	Similarly, the expected strength of transitive competition is:
	\begin{equation} \label{eqn: independent edges null model transitive}
	\mathbb{E}[||F_t||^2] = ||\bar{f}_t||^2 + (m-1) \sum_{k=1}^{E} \left(\frac{\left(P_t\right)_{kk}}{m-1} \right) \sigma_k^2
	\end{equation}
	and the expected total strength of competition is:
	\begin{equation} \label{eqn: independent edges null model}
	\mathbb{E}[||F||^2] = ||\bar{f}||^2 + E \bar{\sigma}^2
	\end{equation}
	where $\bar{\sigma}^2$ is the average of the variance in the flow on each edge. Equation \ref{eqn: independent edges null model} is valid even if the edges are not independent, as the projector onto the full space is simply the identity.
	
	Equations \ref{eqn: independent edges null model cyclic} - \ref{eqn: independent edges null model} show that the contribution to the expected strength of competition from the variances is not distributed equally between the transitive and cyclic spaces. Instead, the amount that is cyclic is proportional to the number of cycles, while the amount that is transitive is proportional to the number of competitors. As a result, adding edges to a network will typically increase the expected degree to which competition is cyclic. It follows that sparse networks with randomly drawn edge flows will be relatively more transitive than would be expected given $\bar{f}$, while dense networks will typically be more cyclic. It also follows that, for a posterior distribution of possible edge flows given observed data, uncertainty will likely lead to an overestimate of the degree to which competition is cyclic, if the network is dense. 
	
	Further simplifications emerge when a network is edge-transitive or has homogeneous variances $\sigma_k^2.$ A network is edge-transitive if the edges are indistinguishable once the node labels are removed. This symmetry implies that $p_{kk}$ is  independent of $k$, regardless the space the projector maps onto. 
	Therefore $\left(P_c\right)_{kk}=L/E$ and $\left(P_t\right)_{kk} = (m-1)/E$.
    Thus:
	\begin{equation} \label{eqn: independent edges symmetric graph null model}
	\begin{aligned}
	& \mathbb{E}[||F_c||^2] = ||\bar{f}_c||^2 + L \bar{\sigma}^2  \\
	& \mathbb{E}[||F_t||^2] = ||\bar{f}_t||^2 + (m-1) \bar{\sigma}^2  \\
	& \mathbb{E}[||F||^2] = ||\bar{f}||^2 + E \bar{\sigma}^2 
	\end{aligned}
	\end{equation}
	where $\bar{\sigma}^2 = \frac{1}{E} \sum_{k=1}^E \sigma_k^2$.
	
	Any symmetric network, or complete network, is edge-transitive, so these equations apply to all symmetric networks and all complete networks with edge flows drawn independently on each edge. Alternatively, if  the variances $\sigma_k^2$ do not depend on $k$, then any weighted average of the variances is equal to $\bar{\sigma}^2$. In this case equations \ref{eqn: independent edges symmetric graph null model} also apply.

	These results show that, in general, the expected strengths of cyclic and transitive competition depend on the expected edge flow, the uncertainty in the edge flow, and the topology of the network. 
	Increasing the uncertainty in the edge flow increases the expected strength of both cyclic and transitive competition, but does not increase both equally. 
	If the graph is sparse, then increasing the uncertainty will typically promote transitive competition more than cyclic. 
	If the graph is dense, then increasing the uncertainty will typically promote cyclic competition more than transitive. If a tournament is complete, then $E = m(m-1)/2$ so $(m-1)/E = 2/m$ and $L/E = 1 - 2/m$. It follows that for a complete tournament with more than four competitors, any uncertainty in the edge flow will typically bias competition to appear more cyclic than transitive. This is necessarily true if the edges are drawn independently, and the graph is either edge-transitive or the variances on each edge are all the same.
	
	Numerical studies have suggested that filling in missing edges with randomly drawn $F$  typically overestimates the degree to which competition is cyclic \cite{Shizuka}. 
	Our result provides a rigorous explanation for this observation.
	When the edge flow $F$ is drawn randomly to fill in missing data, it is usually drawn independently and identically distributed, cf.~\cite{de_Vries}. 
	From equation \ref{eqn: independent edges symmetric graph null model} it is clear that if edges are added until the network is complete, then, for any tournament with more than four competitors, the resulting ``imputed" tournament will likely be significantly  more cyclic than the original tournament. 
	Therefore, unless the edge flows are well-modeled by assuming that the $F_k$ are independent and identically distributed, \textit{and} that all pairs of competitors could compete, this procedure is not valid for estimating the strength of cyclic competition in a partially observed tournament.
	
	The simplified equations \ref{eqn: independent edges null model cyclic} - \ref{eqn: independent edges symmetric graph null model} are valid only if the edge flows are drawn independently, which is rarely the case for real-world tournaments. 
	When the edge flows are not drawn independently, the edge flow covariance matrix is not diagonal, and the simplification leading from equation \ref{eqn: generic null model} to equation \ref{eqn: independent edges null model cyclic} no longer holds.  This makes it more challenging to identify how the topology of the network promotes or suppresses cyclic competition. 
	Nevertheless, as we show in the next section, using a more principled model for sampling $F$, ensures that the covariance matrix $V$ takes on a canonical form. This form clarifies the interaction between the topology of the network and the distribution of edge flows.

	\subsection{Trait-Performance} \label{sec: Trait Performance Models}
		
	% discuss null model with correlations from performance function
	
	The outcomes of real-world competition events are typically influenced by a constellation of  underlying traits of the competitors.  
	Examples of trait-based competition models abound, ranging from sports\footnote{Some predictive tennis models estimate the probability that one competitor will beat another based on a parameterized model for the probability that each player will win a point, where the underlying parameters depend on traits of the players \cite{Kovalchik}. Similarly, considerable effort has been devoted to predictive models for baseball based on team and player statistics \cite{Soto}.} to biology.
	\footnote{Ecological studies of competition for dominance in social hierarchies have analyzed how traits confer success, because selection acts on heritable traits contributing to reproductive success. 
	Examples include competition among male northern elephant seals \cite{Haley} and male Cape dwarf chameleons \cite{Stuart}. 
	Relevant traits for elephant seals include body mass, length, age, and time of arrival on the beach \cite{Haley}. Relevant traits for chameleons include body mass, length from snout to base of tail, length of the tail, jaw length, head width, casque size, and size of a pink colored flank patch used in signaling \cite{Stuart}.}  
	In some cases, trade-offs inherent in certain traits have been observed to lead to cyclic competition between organisms \cite{Kerr,Sinervo}.\footnote{Two particularly famous examples are side-blotched lizards and colicin producing \textit{E.~coli} \cite{Kerr,Sinervo}. 
	In the former example, large orange-throated males maintain large territories, medium blue-throated males defend small territories, while small yellow-throated `sneaker' males resemble females and do not maintain territories. 
	Orange-throated males typically defeat the smaller blue-throated males, who defeat the even smaller yellow throated males, who defeat the orange throated males by sneaking into their territories \cite{Sinervo}. 
	In the latter example, three strains of \textit{E.~coli} were grown in direct competition in a laboratory setting. 
	The first strain produced a colicin toxin, the second was susceptible to the toxin, and the third was resistant to the toxin but not toxin-producing. 
	In the absence of the resistant strain, the toxic strain could outcompete the susceptible strain. 
	In the absence of the toxic strain, the susceptible strain could outcompete the resistant strain, which reproduced more slowly because resistance is costly. But, in the absence of the susceptible strain, the resistant strain could outcompete the toxic strain by reproducing more quickly \cite{Kerr}.} 
	In such examples, trade-offs lead to advantages against certain opponents, and weaknesses that are exploited by others. 
In evolutionary biology, trade-offs of this kind challenge the notion that members of intransitive communities can be consistently ranked according to fitness. 
Intransitivity can lead to deeply counterintuitive evolutionary dynamics \cite{Frean, Johnson}, and may promote biodiversity since no single species has an absolute advantage over all competitors \cite{Reichenbach_a,Reichenbach_b,Reichenbach_c,Reichenbach_d,Soliveres}. 
These considerations motivate a study of how demographics (the distribution of traits), and the way traits confer success, either promote or suppress cyclic competition.

	Therefore, we now suppose that win probabilities $p$ can be modeled as a function of some underlying traits $x$ of each competitor. 
	Let $X(i) = [X_1(i),\ldots,X_T(i)]$ denote the $T$ randomly sampled traits of the $i^{th}$ competitor. Then let $ f(x,y)$ be a performance function, such that $f(x,y)$ is the log-odds that a competitor with traits $x$  would beat a competitor with traits $y$.
	
	To construct a trait-performance model assume that:
	
	\begin{enumerate}
	\item The trait vectors of the competitors are drawn independently and identically from a trait distribution $\pi_x$. 
	
	\item There exists a performance function $f(x,y)$ that maps from $\mathbb{R}^T \times \mathbb{R}^T$ to $\mathbb{R}$. We require that the performance function is alternating $f(x,y) = - f(y,x)$ for any trait vectors $x$ and $y$ in the support of $\pi_x$. This ensures that $f$ can be used to generate an  edge flow. It also ensures that the performance function is fair, $\mathbb{E}[f(X,Y)] = 0$, since if $X$ and $Y$ are drawn i.i.d then $\mathbb{E}[f(X,Y)] = \mathbb{E}[f(Y,X)] = -\mathbb{E}[f(X,Y)]$ which implies $\mathbb{E}[f(X,Y)] = 0$.
	
	\item There exists a connected competitive network $\mathcal{G}_{\rightleftarrows}$ with edges representing possible competition events, and the network is either fixed a priori or sampled independently from the traits.
	\end{enumerate}
	
	Assumptions 1 and 3 are the most restrictive. The first assumes all competitors are drawn from the same demographic pool. Different demographic pools can be incorporated into the model by adding a trait which indexes which pool each competitor is sampled from, provided that trait can be sampled independently of the graph. For example, Major League Baseball team budgets vary widely. In 2018 the Yankees' total value was over 4.6 billion dollars, which was more than the total value of the bottom six teams combined \cite{Ozanian}. This difference resoucres gives high value teams the opportunity to pay higher salaries\footnote{For example, in 2019 the Yankees' combined payroll  was three times larger than the Marlins'.} and thus attract star players. Thus rich teams are in a different demographic pool than poor teams, so the wealth of the teams could be incorporated as one of their traits.
	
	The third assumption treats the network topology (who competes with whom) independently from the traits of the competitors. This may not be realistic if competitors avoid competing when they are likely to lose \cite{Sismanis}. This also limits our ability to model systems where traits are heritable, or distributed differently across different clusters of competitors (different divisions, or local populations).
	
	The second assumption is the least restrictive since it is valid whenever the probability that one competitor beats another can be conditioned on the traits of the competitors, independent of their location on the network.
	
	Under these assumptions, we define a trait-performance model as follows. First, sample $X(i) \sim \pi_x$ for all competitors $i$. Then, set $F_{k} = f(X(i(k)),X(j(k)))$, where $i(k),j(k)$ are the endpoints of edge $k$.

%%	Given a trait-performance model the expected strength of transitive and cyclic competition can computed analytically - and depends monotonically on the correlation coefficient between $f(X(i),X(j))$ and $f(X(i),X(k))$ for $k \neq j$. In general, the larger this correlation the more transitive competition is expected to be. The value of this correlation coefficient depends on the form of the performance function, and the distribution of traits, while the expected strength of transitive and cyclic competition are functions of the correlation coefficient, the variance in $f(X,Y)$, and the dimensions $m, E, L$. Therefore, the expected degree to which competition is transitive or cyclic can be understood directly from the correlation structure of the edge flow - which could be computed analytically for some special cases, or numerically for more realistic models.
	
	\begin{snugshade}
	\textbf{Theorem 7: (Trait-Performance)} \label{thm: expected transitivity and correlation coefficient}
		 Let $\mathcal{G}_\rightleftarrows$ be a competitive network satisfying assumption $3$.
		 If the traits of each competitor are drawn independently from $\pi_x$, and the edge flow is defined by $F_k = f(X(i(k)),X(j(k)))$ where $f(x,y)$ is an alternating performance function, then the covariance $V$ of the edge flow has the form:
		\begin{equation} \label{eqn: generic covariance}
		V = \sigma^2 \left[I + \rho \left(G G^T - 2 I \right) \right]
		\end{equation}
		where $\sigma^2$ is the variance in $F_k$ for arbitrary $k$, and $\rho$ is the correlation coefficient between $f(X,Y)$ and $f(X,W)$ for $X,Y,W$  drawn i.i.d from $\pi_x$.
		
		Moreover:
		\begin{equation}
		\mathbb{E} \left[\frac{1}{E}||F||^2 \right] = \sigma^2  \xrightarrow{\text{decompose}}
		\left\{
		\begin{aligned}
		& \mathbb{E} \left[\frac{1}{E}||F_t||^2 \right] = \sigma^2 \left[\frac{(m-1)}{E} + 2 \rho \frac{L}{E} \right] \\
		& \mathbb{E} \left[\frac{1}{E}||F_c||^2 \right] = \sigma^2 \left(1 - 2 \rho \right) \frac{L}{E} \end{aligned} \right.
		\end{equation}
		
		Therefore, the expected absolute strength of competition is independent of $\rho$, the size of the transitive component is monotonically increasing in $\rho$, and the size of the cyclic component is monotonically decreasing in $\rho$. The correlation $\rho$ ranges from 0 to $1/2$, and if $\rho = 1/2$ then competition is perfectly transitive.
		
	\end{snugshade}
	
	\textbf{Proof: } First consider the covariance matrix $V$. 
	
	Since the trait vectors are drawn i.i.d from the trait distribution, the diagonal entries of the covariance are given by:
	\begin{equation}
	V_{kk} = \mathbb{E}\left[ \left(f(X(i(k),X(j(k)) \right)^2 \right] = \mathbb{E}\left[ \left(f(X,Y) \right)^2 \right] \equiv \sigma^2
	\end{equation}
	where $X,Y$ are drawn i.i.d from the trait distribution, and $\sigma^2$ is the variance in $f(X,Y)$. 
	Thus, the diagonal entries of the covariance are identical.
	
	The off-diagonal entries are $V_{kl} = \mathbb{E}\left[ f(X(i(k)),X(j(k)))\cdot f(X(i(l)),X(j(l)))  \right].$
	
	Suppose the edges $k$ and $l$  do not share an endpoint. Then $i(k) \neq i(l)$ or $j(l)$ and $j(k) \neq i(l)$ or $j(l)$. Then $f(X(i(k)),X(j(k)))$ is a function of two random vectors, and $f(X(i(l)),X(j(l)))$ is a function of two other random vectors, where the pair of random vectors are independent. It follows that $f(X(i(k)),X(j(k)))$ is independent of $f(X(i(k)),X(j(k)))$. Then, since competition is fair for all alternating performance functions $V_{kl} = \mathbb{E}\left[ f(X(i(k)),X(j(k)))\cdot f(X(i(l)),X(j(l)))  \right] = \mathbb{E}\left[ f(X(i(k)),X(j(k)))\right] \mathbb{E}\left[ f(X(i(l)),X(j(l)))  \right] = 0$. It follows that the support of the covariance matches the adjacency structure of the edges of the competition network.
	
	If the edges do share an endpoint, then there are four possibilities. Either $i(k) = i(l)$, $j(k) = j(l)$, $i(k) = j(l)$, or $j(k) = i(l)$. 
	We say that the edges are \emph{consistently oriented} if they share either the same starting point or the same ending point, and are \emph{inconsistently oriented} if the endpoint of one is the start of another. 
	Since all the trait vectors are drawn i.i.d., we suppress the indices and let the three trait vectors $Y,W,Z$ be drawn i.i.d. from $\pi_x$. 
	The performance function is alternating, so:
	\begin{equation}\label{eq4.10}
	\begin{aligned}
	& \mathbb{E}[f(Y,W) f(Y,Z)] = \mathbb{E}[f(W,Y)f(Z,Y)] \equiv \rho \sigma^2 \\
	& \mathbb{E}[f(Y,W) f(Z,Y)] = \mathbb{E}[f(W,Y)f(Y,Z)] = -\mathbb{E}[f(Y,W)f(Y,Z)] = -\rho \sigma^2
	\end{aligned}
	\end{equation}
	where $\rho$ is the correlation coefficient between $f(Y,W)$ and $f(Y,Z)$. Notice that a positive correlation indicates that the probability that $A$ beats $B$ is increased by conditioning on the event that $A$ beats $C$. 
	
	\begin{figure}[t] \label{fig: edge graph}
		\centering
		\includegraphics[scale=0.3]{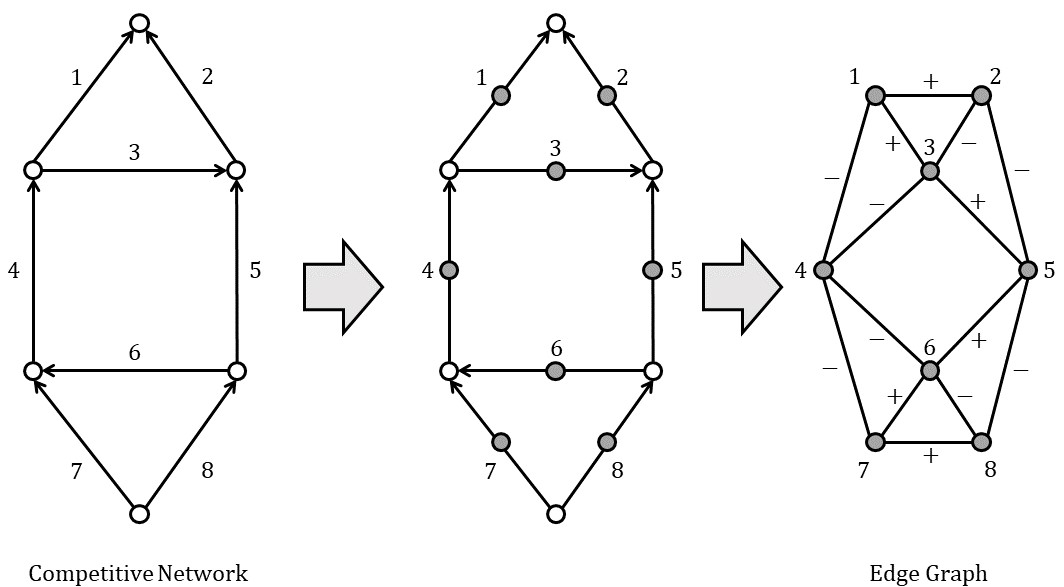}
		\caption{The edge graph (right) associated with a competitive network (left). The middle panel shows an intermediate graph where a node has been introduced for each edge. The edges of the competitive network become the nodes of the edge graph. The edges of the edge graph correspond to nodes in the competitive network that are the shared endpoint of a pair of edges. These are labelled with a $+$ or $-$ to indicate whether the edges are consistently or inconsistently oriented with respect to the shared endpoint.}
	\end{figure}
	
	The edge graph is the graph with a node for each edge in the competition network, and with an undirected edge between nodes corresponding to connected edges in the competition network (Figure \ref{fig: edge graph}).
	Let $A_E$ be the weighted adjacency matrix for the edge graph with ${a_{E}}_{kl} = +1$ or $-1$ if edges $k$ and $l$ are consistently or inconsistently oriented with respect to a shared endpoint. Then:
	\begin{equation}
	V = \sigma^2 \left[I + \rho A_E \right].
	\end{equation}
	
	The weighted adjacency matrix $A_E$ for the edge graph is equal to $G G^T - 2 I$ since:
	\begin{equation} \label{eqn: edge adjacency}
	[G G^T]_{kl}  = (e_{i(k)} - e_{j(k)})^T(e_{i(l)} - e_{j(l)}) = \left\{ \begin{aligned} & 2 \text{ if } k = l \\ & 1 \text{ if } i(k) = i(l) \text{ or } j(k) = j(l) \\ & -1 \text{ if } i(k) = j(l) \text{ or } j(k) = i(l) \\ & 0 \text{ else } \end{aligned} \right\} 
	\end{equation}
	
	\noindent where $e_i \in \mathbb{R}^m$ is the indicator vector for node $i$.  Thus we establish  equation \ref{eqn: generic covariance}.
	
	All of the absolute measures of the strength of competition (squared) are given by the squared length of the orthogonal projection of the edge flow onto some subspace. Let $P_S$ be an arbitrary orthogonal projector onto some subspace $S$. By construction, the edge flow is zero mean, therefore, by equation \ref{eqn: generic null model}, the expected value of the associated measure is:
	\begin{equation} \label{eqn: matrix inner product}
	\mathbb{E}\left[||F_S||^2 \right] = \text{trace}(P_S V)
	\end{equation}
	where $V$ is the covariance matrix of the edge flow $F$. 
	
	The intensity of competition, $||F||^2$, corresponds to the projector $I$, $||F_t||^2$ corresponds to the projector $P_t$, and $||F_c||^2$ corresponds to the projector $P_c$. Then, by equation \ref{eqn: matrix inner product}:
	\begin{equation}
	\mathbb{E} \left[\frac{1}{E}||F||^2 \right] = \frac{1}{E} \text{trace}(V) = \frac{E}{E} \sigma^2 = \sigma^2.
	\end{equation}
	
	This formula establishes that the absolute strength of competition only depends on the variance $\sigma^2$ in each individual performance function. 
	
	To compute $||F_t||^2$, use equation \ref{eqn: matrix inner product} with projector $P_t$:
	\begin{equation}
	\begin{aligned}
	\mathbb{E} \left[\frac{1}{E}||F_t||^2 \right] & = \frac{1}{E} \text{trace}(P_t V) = \frac{\sigma^2}{E} \text{trace}\left(P_t[I + \rho(G G^T - 2 I)] \right) \\
	& = \frac{\sigma^2}{E} \text{trace}\left(P_t \right) + \frac{\rho \sigma^2}{E} \text{trace}\left(P_t (G G^T) \right) -  \frac{2 \rho \sigma^2}{E} \text{trace}\left(P_t \right).
	\end{aligned}
	\end{equation}
	
	The trace of an orthogonal projector equals the dimension of the subspace it projects onto, so $\text{trace}(P_t) = m-1$. The range of $G G^T$ is in the range of $G$, which is the subspace $P_t$ projects onto. It follows that $P_t G G^T = G G^T$ so $\text{trace}(P_t G G^T) = \text{trace}(G G^T) = 2 E$ (see equation \ref{eqn: edge adjacency}). Therefore:
	\begin{equation}
	\mathbb{E}\left[\frac{1}{E}||F_t||^2 \right] = \sigma^2 \left[ \frac{m-1}{E} + 2 \rho \frac{E - (m-1)}{E} \right] = \sigma^2 \left[ \frac{m-1}{E} + 2 \rho \frac{L}{E} \right].
	\end{equation}
	
	Since $L \geq 0$,  $\mathbb{E}[\frac{1}{E}||F_t||^2]$ increases monotonically in $\rho$: the larger $\rho$, the more $A$ beating $B$ is correlated with $A$ beating $C$,  implying transitive competition.
	
	To compute the expected absolute strength of cyclic competition (squared) we take advantage of the orthogonality of the decomposition $f = f_c + f_t $:
	\begin{equation}
	\mathbb{E} \left[\frac{1}{E}||F_c||^2 \right] = \mathbb{E} \left[\frac{1}{E}||F||^2 \right] - \mathbb{E} \left[\frac{1}{E}||F_t||^2 \right] = \sigma^2 \left[1 - 2 \rho \right] \frac{L}{E}.
	\end{equation}
	
	It follows that the expected absolute strength of cyclic competition is monotonically decreasing in the correlation coefficient $\rho$. Note that, as when considering the generic null models, dense networks promote cyclic competition.
	
	To conclude we show that $\rho \in [0,1/2]$, so the expected measures are maximized and minimized when $\rho$ is 0 or 1/2, respectively.
	
	The correlation $\rho$ is nonnegative since $W$ and $Z$ are i.i.d., thus $f(y,W)$ and $f(y,Z)$ are also i.i.d., so:
	\begin{equation} \label{eqn: rho nonnegative}
	\begin{aligned}
	& \sigma^2 \rho = \mathbb{E}_{Y,W,Z}[f(Y,W)f(Y,Z)]  = \int_{\mathbb{R}^T} \mathbb{E}_{W,Z}[f(y,W)f(y,Z)] \pi_x(y) dy \\
	&  =  \int_{\mathbb{R}^T} \mathbb{E}_W[f(y,W)] \mathbb{E}_Z[f(y,Z)] \pi_x(y) dy  = \int_{\mathbb{R}^T} \mathbb{E}_{W}[f(y,W)]^2  \pi_x(y) dy \geq 0
	\end{aligned}
	\end{equation}
	Here expectation is taken with respect to the variables in the subscript. 
	
	To prove that $\rho \leq 1/2$, note that all covariance matrices are positive semi-definite, so, for any vector $u$:
	\begin{equation}
	u^T V u = \sigma^2 u^T(I + \rho (G G^T - 2 I)) u = \sigma^2 (1 - 2 \rho)||u||^2 + \rho u^T G G^T u \geq 0.
	\end{equation}
	
	If $E > m-1$, then the network has at least one loop, so the range of $C^T$ is non-empty, hence the null-space of $G^T$ is non-empty. Choosing $u$ perfectly cyclic sets $G^T u = 0$ so $\sigma^2 (1 - 2 \rho)||u||^2 \geq 0$ which requires $\rho \leq \frac{1}{2}$. If $E = m-1$ then the network is a tree, so all competition is necessarily perfectly transitive.
	
	It follows that the expected absolute strength of \textit{transitive} competition is minimized when $\rho = 0$, and maximized when $\rho = 1/2$. In contrast, the expected strength of \textit{cyclic} competition is maximized when $\rho = 0$, and minimized when $\rho = 1/2$. 
	
	If $\rho = 1/2$ then $\mathbb{E}[||F_c||^2] = 0$. The measure is  nonnegative for all edge flows. Therefore, its expected value is only zero if the probability that $||F_c||^2 \neq 0$ is zero. In this case, the tournament is arbitrage free. It follows that, if $\rho = 1/2$, then the tournament must be perfectly transitive.\footnote{Note that $\rho = 1/2$ guarantee perfect transitivity but $\rho = 0$ does not guarantee that the tournament is perfectly cyclic. A counterexample suffices to explain why. Suppose each competitor chooses rock, paper, or scissors uniformly and independently. Suppose there are three competitors and the tournament is complete. Then, in order for the tournament to be perfectly cyclic, rock must be chosen by one competitor, scissors by another, and paper by the last. There are 6 ways this can happen but there are 27 possible tournaments. Therefore a three competitor system has a 21/27 chance of being perfectly transitive, even when the underlying performance function is clearly cyclic.  } $\blacksquare$
	
	% give formulas for sigma and rho
	
	Theorem \ref{thm: expected transitivity and correlation coefficient} establishes that the expected degree to which competition is transitive or cyclic depends principally on the density of the network, and the correlation structure of $F$. In particular, the degree to which a network is cyclic or transitive depends on the correlation between the performance of $A$ against $B$ with the performance of $A$ against $C$. The larger this correlation, the more consistently each competitor performs, hence the more consistent the network is with a set of ratings. 
	
	The variance $\sigma^2$ and the correlation coefficient $\rho$ could be computed given an assumed trait distribution $\pi_x$ and performance function $f(x,y)$. This could be done analytically if $\pi_x$ and $f$ lead to simple calculations. Otherwise, $\sigma^2$ and $\rho$ can be approximated numerically. The analytic method follows.
	
	Suppose that $X,Y$ are drawn from a sample space $\Omega$ which is a subset of $\mathbb{R}^T$. Then, for trait distribution $\pi_x$, the variance in $f(X,Y)$ is given by $\sigma^2 = \mathbb{E}_{X,Y}\left[f(X,Y)^2 \right] = \int_{\Omega} \int_{\Omega} f(x,y)^2 \pi_x(y) \pi_x(x) dy dx.$ Then, substituting into equation \ref{eqn: rho nonnegative}:
	\begin{equation} \label{eqn: rho as an integral}
	\rho = \frac{\int_{\Omega} \left(\int_{\Omega} f(x,y) \pi_x(y) dy \right)^2 \pi_x(x) dx}{\int_{\Omega} \int_{\Omega} f(x,y)^2 \pi_x(y) \pi_x(x) dy dx}.
	\end{equation}
	
	Note that the correlation coefficient is only large if it is possible to find some set of traits which are expected to perform either well or poorly on average, and if these traits occur with sufficient probability. That is, there must be some $x$ such that $|\mathbb{E}_Y[f(x,Y)]|$ is large, and such that  $\pi_x(x)$ is not too small. From this expression, it is not surprising that the expected strength of transitive competition is monotonically increasing in $\rho$. If there is a set of traits $x$ which, on average, either overperform or underperform against randomly drawn opponents, and are frequently sampled, then a random sample of $m$ competitors is expected to include some who perform well, and some poorly, against their neighbors. If, on the other hand, the expected performance conditioned on traits $x$ is close to neutral, then $\rho$ is small and competition is expected to be cyclic. 
	In a rock-paper-scissors style game in which competitors are randomly and uniformly assigned  rock, paper, or scissors, then conditioning on receiving a particular trait does not change the probability that an individual with that trait will win most contests, hence the tournament is expected to be highly cyclic.
	
	Another way to read equation \ref{eqn: rho as an integral} is as follows. Define the expected performance of traits $x$ to be $\mathbb{E}_Y[f(x,Y)]$. Then since $\mathbb{E}_{X}[\mathbb{E}_Y[f(X,Y)]] = \mathbb{E}_{X,Y}[f(X,Y)] = 0$, $\mathbb{E}_{X}[\mathbb{E}_Y[f(X,Y)]^2]$ is the variance in the expected performance. Therefore $\rho$ is the ratio of the variance in the expected performance to the variance in performance. 
	A large variance in the expected performance means we are likely to sample some competitors who perform well, or poorly, against most opponents. Consequently, the sampled edge flow is expected to be more transitive than cyclic.
	
    Rereading Theorem \ref{thm: expected transitivity and correlation coefficient} in this way leads to the following insight:
    
    \begin{snugshade}
    \textbf{Corollary 8: } \label{Corollary: uncertainty in expectation} If the traits $W,X,Y$ are sampled independently from $\pi_x$ and $F = f(X,Y)$ then the correlation coefficient $\rho$ is proportional to the variance in the expected performance:
    \begin{equation} \label{eqn: rho is variance in expectation}
        \rho = \frac{1}{\sigma^2} {\rm{cov}}(f(X,Y),f(X,W)) =  \frac{1}{\sigma^2}{\rm{Var}} \left( \mathbb{E}[F|X] \right).
    \end{equation}
    
    Let $\nu$ be the expected variance in the  performance:
    \begin{equation} \label{eqn: nu is expected variance}
        \nu =  \frac{1}{\sigma^2} \mathbb{E} \left[ {\rm{Var}}(F|X) \right].
    \end{equation}
    
    Then $\nu = 1 - \rho$, so $\mathbb{E}[||F_c||^2]$ is monotonically increasing in $\nu$, $\mathbb{E}[||F_t||^2]$ is monotonically decreasing in $\nu$, and $\nu = \frac{1}{\sigma^2} {\rm{Var}}[f(X,Y) - f(X,W)]$.
    
    \end{snugshade} 
    
    \textbf{Proof: } The proof of equation \ref{eqn: rho is variance in expectation} is given by equation \ref{eqn: rho as an integral}, and the fact that $\mathbb{E}[F] = 0$. Then $\nu = 1 - \rho$ follows by the law of total variance:
    \begin{equation}
    \begin{aligned}
        \sigma^2 & = \text{Var}(F) = \mathbb{E} \left[ \text{Var}(F|X) \right] + \text{Var} \left[ \mathbb{E}(F|X) \right]  = \sigma^2(\rho + \nu).
    \end{aligned}
    \end{equation}
    
    Since $\mathbb{E}[||F_c||^2]$ is decreasing in $\rho$, it is increasing in $\nu$. Similarly, since $\mathbb{E}[||F_t||^2]$ is increasing in $\rho$, it is decreasing in $\nu$.
    
    The final expression for $\nu$ follows from $\sigma^2 \nu = \sigma^2 (1 - \rho) = \text{Var}[f(X,Y)] - \text{cov}[f(X,Y),f(X,W)]$. Since $Y$ and $W$ are i.i.d., $\text{Var}[f(X,Y)] = \frac{1}{2} (\text{Var}[f(X,Y)] + \text{Var}[f(X,W)])$. Substituting in gives $\sigma^2 \nu = \frac{1}{2} \mathbb{E}[(f(X,Y) - f(X,W))^2]$. Since $\mathbb{E}[f(X,Y)]$ equals $\mathbb{E}[f(X,W)]$ this raw second moment is the variance in $f(X,Y) - f(X,W)$.
    $\blacksquare$

   \begin{figure}[t] \label{fig: trait performance bar schematic}
		\centering
		\includegraphics[scale=0.33]{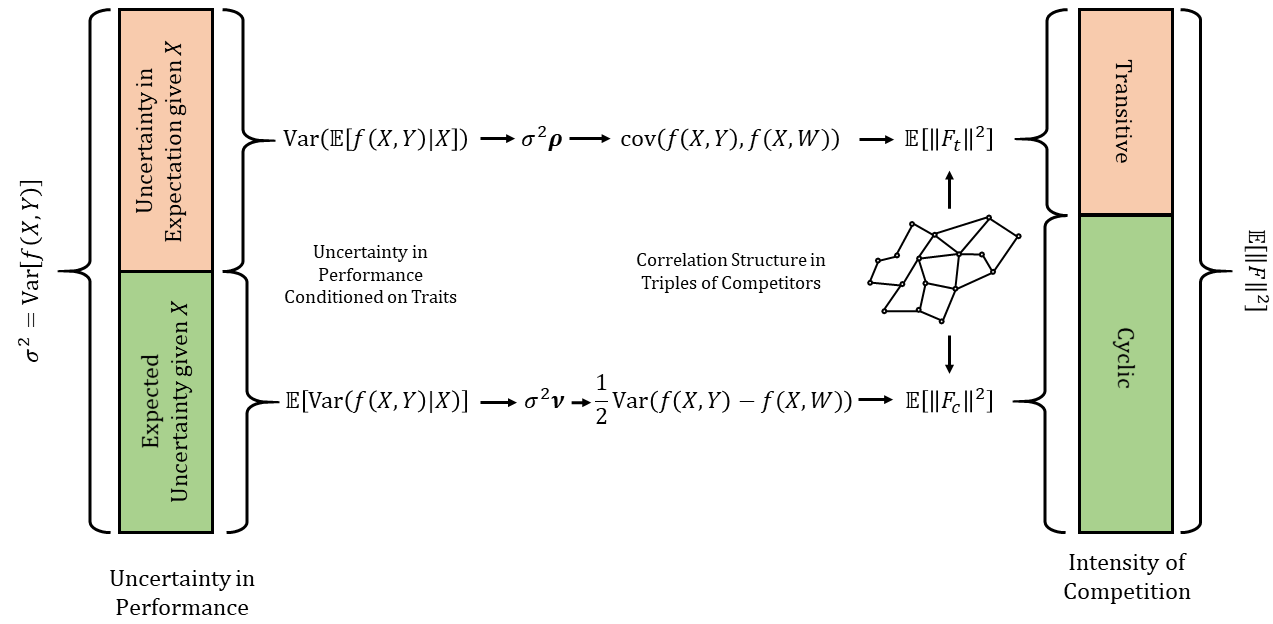}
		\caption{ A schematic representing the conclusions of Theorem \ref{thm: expected transitivity and correlation coefficient} and Corollary \ref{Corollary: uncertainty in expectation}. The left hand side decomposes the uncertainty in performance into the uncertainty in the expected performance given $X$, and the expected uncertainty in the performance given $X$. 
		These uncertainties are converted into $\rho$ and $\nu$ which describe the correlation structure of triples of competitors. The sizes of $\rho$ and $\nu$, plus the topology of the network, determine the expected sizes of the transitive and cyclic components. Thus we convert a decomposition of the uncertainty in the performance into a decomposition of the intensity of the edge flow representing competition. }
	\end{figure}
    
    Theorem \ref{thm: expected transitivity and correlation coefficient} identifies which statistical feature of the trait distribution and performance function promotes transitive and suppresses cyclic competition. Corollary \ref{Corollary: uncertainty in expectation} complements this understanding by showing which feature suppresses transitive and promotes cyclic competition. 
    Transitive competition is promoted by the uncertainty in expected performance, $\text{Var}[\mathbb{E}(F|X)]$, and suppressed by the expected uncertainty, $\mathbb{E}[\text{Var}(F|X)]$. Conversely,
    cyclic competition is suppressed by uncertainty in the expected performance, and promoted by expected uncertainty. 
    If the uncertainty in expected performance is large, then we are likely to sample some competitors who are consistently better, or worse, than their neighbors, hence competition is mostly transitive. 
    If the expected uncertainty in performance is large, then it is difficult to predict the performance of a single competitor against their neighbors, since performance is competitor dependent, hence competition is mostly cyclic. \
    
    Together Theorem \ref{thm: expected transitivity and correlation coefficient} and Corollary \ref{Corollary: uncertainty in expectation} provide conceptual bridges between uncertainty in the flow on each edge, correlation structure on edges that share an endpoint, and cyclic/transitive structure on the  network (see Figure \ref{fig: trait performance bar schematic}). They establish the intuitive statements that conclude the introduction (p.~\pageref{stm:1b}). For example, the expected uncertainty in the performance of $A$ against a random competitor is $\sigma^2 \nu = \frac{1}{2}\mathbb{E}_X[\text{Var}_Y(f(X,Y)|X)]$. Thus, \textit{``the less predictable the performance of $A$ against a randomly drawn competitor, the more cyclic the tournament''} (see \labelcref{stm:1b}). Then, by the equivalence of $\mathbb{E}_X[\text{Var}_Y(f(X,Y)|X)]$ to $\text{Var}(f(X,Y) - f(X,W))$, \textit{``the more the performance of $A$ depends on their opponent, the more cyclic the tournament.''}

	It remains to understand how the choice of trait dimension, trait distribution, and performance function influence $\rho$, and consequently the expected degree of cyclic competition. 
	We provide an illustrative example  below. 
	
	\section{Example} \label{sec: Case Studies}
	
	Suppose that each competitor has a set of $T$ traits. 
	Assume that the traits are chosen so that the performance function $f(x,y)$ is non-decreasing in  $x_j$, and non-increasing in $y_j$, for all $j$. 
	This amounts to choosing a sign convention for each trait so that increasing any trait improves performance. 
	Then a competitor with traits $x$ has an advantage (in trait $j$) over an opponent with traits $y$ if $x_j > y_j$. 
	
	In some events, competitors with a large advantage in a given trait can dominate, so that the event is primarily mediated by that trait. 
	That is, competitors press their advantages. 
	For example, a performance function of this type is the extremal performance function $f(x,y) = x_j - y_j$, where $j$ is the dimension in which this difference is largest in magnitude,  $j=\text{argmax}_j|x_j-y_j|$. 
	In the extremal performance model, the performance is completely controlled by the largest advantage, so competitive events are as one-sided as possible, given the competitor's traits.
	
	Consider, in contrast, a competitive event in which competitors cannot press their advantages. For example: $f(x,y) = x_j - y_j$ for the dimension $j=\text{argmin}_j|x_j-y_j|$ that minimizes the advantage. 
	This rule could model a contest in which competitors are required to reach a consensus about how to compete in advance or, where the weaker competitor controls which traits primarily mediate the competitive event. Competitors could be motivated or compelled to compete without pressing advantages by an external mediating body. 
	For example, a sports league is motivated to keep teams evenly matched, even if the individual teams are motivated to win. 
	
	Suppose that the traits are drawn i.i.d from either an exponential, Gaussian, or uniform distribution. 
	In each case, the variance of the trait distribution has no effect on $\rho$ so, without loss of generality, each distribution is chosen to have variance one. 
	
	\begin{figure}[t] \label{fig: press your advantage vs fair fight}
		\centering
		\includegraphics[scale=0.16]{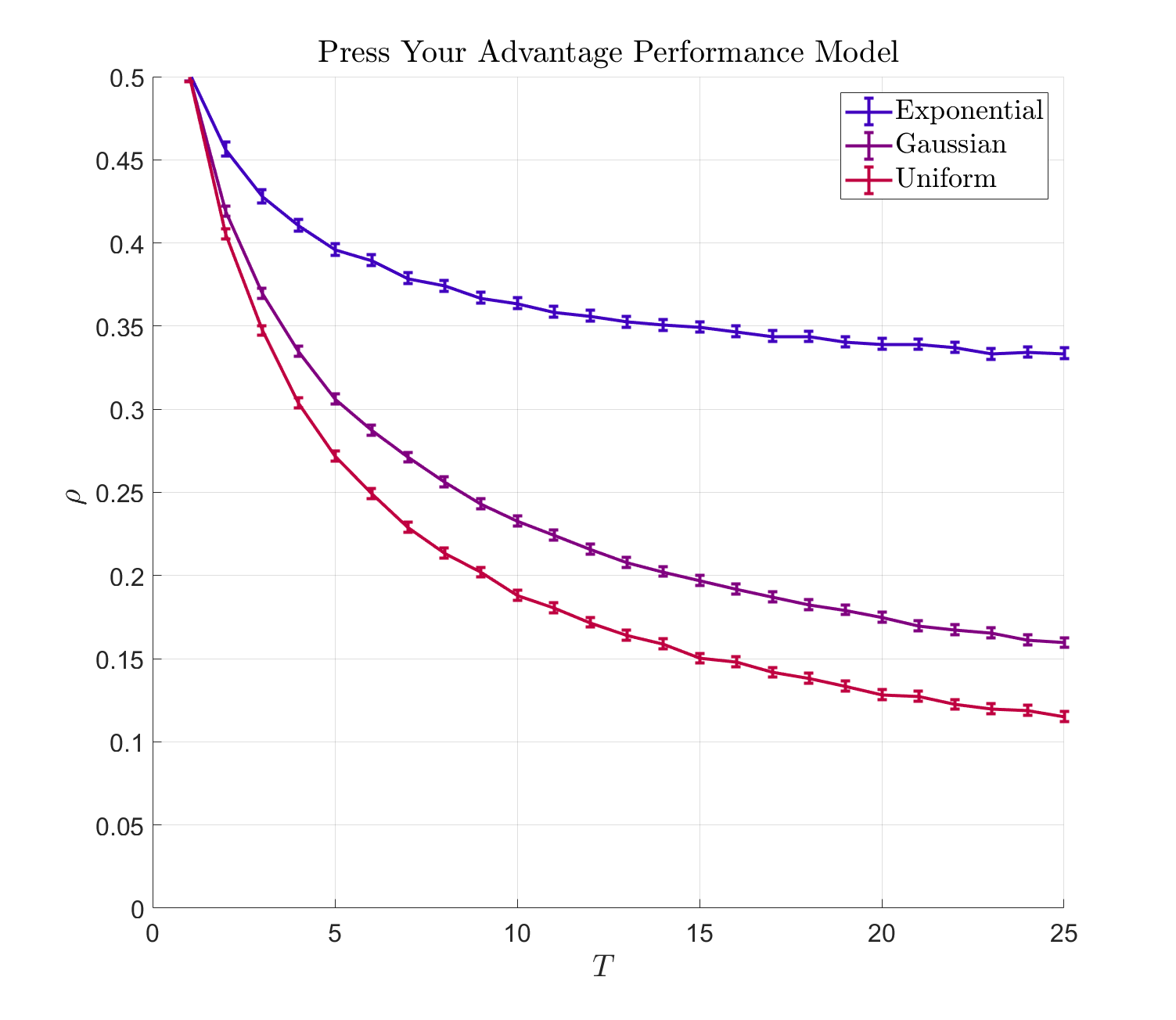}
		\includegraphics[scale=0.16]{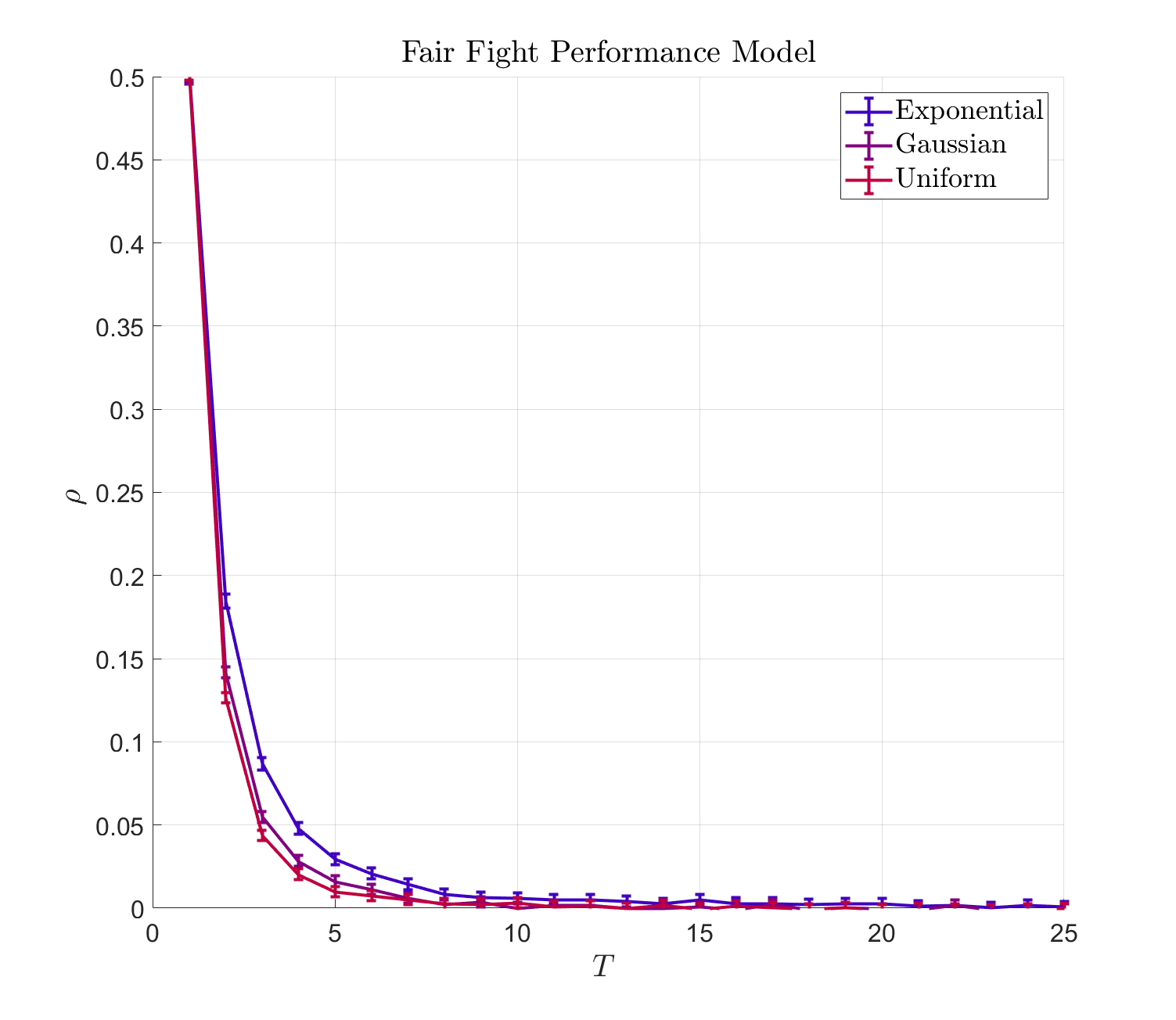}
		\caption{The correlation coefficient $\rho$ for two different performance functions and three different trait distributions as a function of the number of competitive traits. Error bars represent three standard deviations in the estimated correlation coefficient. The ``Press Your Advantage" panel shows $\rho(T)$ for the extremal performance model: $f(x,y) = x_j - y_j$ for $j$ that maximizes the difference. The``Fair Fight" panel shows $\rho(T)$ for $f(x,y) = x_j - y_j$ for $j$ that minimizes the difference. 
		}
		
	\end{figure}
	
	We estimated the correlation coefficient $\rho$ for all six models (two performance functions, three distributions) with trait dimension varying from 1 to 25. To estimate the correlation coefficient for a given model and trait dimension we sampled $10^6$ triples of trait vectors $X,Y,W$ and computed $f(X,Y) f(X,W)$. Averaging over all $10^6$ triples gave an empirical estimate for the covariance, which was then normalized by an empirical estimate of the variance $\sigma^2$. Figure \ref{fig: press your advantage vs fair fight} shows the results. 
	
	For all three choices of trait distribution, $\rho(T)$ was larger if the extremal advantage model was used instead of the fair-fight model. This indicates that, the more competitors can press their advantages, the more transitive competition is, on average. 
	This is not surprising, since in the fair-fight model, the traits mediating performance for competitor $A$ against competitor $B$ are likely different from the traits mediating competition between $A$ and $C$.
	As a result, the success of competitor $A$ is highly competitor dependent. Thus competition is more cyclic. 
	
	Note that this conclusion is much easier to test using the trait-performance theorem (Theorem \ref{thm: expected transitivity and correlation coefficient}) than by sampling a series of random edge flows. 
	Using Theorem \ref{thm: expected transitivity and correlation coefficient}, we only needed to sample trait vectors for triples of competitors to evaluate $\rho$. 
	This simplification greatly reduces the sampling cost. 
	
	In all six models tested, $\rho(T)$ is decreasing in $T$, so the expected proportion of competition that is cyclic is increasing. 
	This matches the results in \cite{Landau_a}, where increasing the trait dimension typically decreased the expected degree of transitivity. This is intuitive, since larger $T$ allows more ways for two competitors to compete, so it is harder to assign a single rating to a competitor.\footnote{Note that while this is often true it is \textit{not} true for all trait-performance models. 
	} 

	When using the extremal performance model the correlation $\rho(T)$ decays much faster in $T$ for Gaussian and uniform traits than for exponential traits. This is because exponentially sampled traits are more likely to include large outliers. Since the extremal performance model sets $f$ to the largest trait difference, the performance is more likely to depend on the outlier traits of each competitor. If a competitor has one particularly large trait, and $T$ is large, then it is unlikely that any other competitor has a comparably large trait value in the same dimension. As a result, the competitor with the largest trait usually competes along that dimension and their performance against other competitors is fairly consistent. This leads to a relatively high $\rho$. 
	
	On the other hand, if the traits are drawn uniformly from $[0,1]$ then no competitor can achieve a universal advantage by having one extremely large trait value. 
	Instead, as the dimension of the trait space increases, competitors succeed by having a large trait value where their opponent has a small trait value - that is, by exploiting their opponents' weaknesses. 
	In this situation, the relevant trait dimension that determines the outcome of competition depends on whom each competitor competes with. Consequently the correlation $\rho$ becomes very small as $T$ becomes large, so competition becomes predominantly cyclic.
	
	In the fair-fight model all three trait distributions produce nearly identical correlations, since outlier traits do not mediate performance. Instead, performance is mediated by average traits, since the smallest advantage $X_j - Y_j$ is likely to come from a trait dimension where both $X_j$ and $Y_j$ are close to their expected values.
	
	This example illustrates the explanatory power of the trait-performance theorem. By separating the influence of network topology from  statistical assumptions about competition, the  theorem
	facilitates numerical hypothesis testing and affords deeper insights by focusing the questions we ask about competitive tournaments.

	\section{Discussion} \label{sec: Discussion}
	
	The discrete HHD  provides a natural, unified method for ranking and measuring intransitivity via a decomposition into perfectly transitive and cyclic components. The expected size of these components can be computed from the correlation structure of the edge flow. 
	Using a trait-performance model simplifies this correlation so that the decomposition of the edge flow can be related to the correlation in adjacent edges and to a decomposition of the uncertainty in the flow. 
	Intuitive statements about the expected sizes of the components can be rigorously proven for such models, which provide conceptual insight, as illustrated in  Section \ref{sec: Case Studies}. Future work should address other case studies, both inspired by real systems and chosen to illustrate generic behavior. 
	
	Further theoretical work could address random network topologies. If the network is sampled independently of the edge flow then the results of Theorem \ref{thm: expected transitivity and correlation coefficient} are largely unchanged.
	Future work might consider random networks whose distribution depends on the traits of each competitor, or ensembles whose traits are not i.i.d. 
	For example, competitors who are neighbors in the network might have positively correlated traits. 
	This would be important in an evolutionary setting where the hereditary nature of traits matters. 
	Future work could also investigate null models with differently structured covariances in the edge flow. Studying these models will help contextualize the HHD when applied to real tournaments.
	
	This work can be extended to data from real tournaments. By studying win-loss records it is possible to infer the log odds edge flow, and thus estimate the components of the HHD. This work provides context by offering comparison to null models. Moreover, when an exhaustive win-loss record is not available, this work suggests that the expected size of the cyclic component could be estimated by estimating the correlation coefficient $\rho$, which may be easier to estimate robustly.

\section{Acknowledgments}
We would like to thank Lek-Heng Lim, Robin Snyder, and Michael Hinczewski for their helpful conversations. We also thank Gilbert Strang for his help revising. This work was funded by NSF grant DEB-1654989.

\bibliographystyle{siamplain}
\bibliography{Refs}

\end{document}